\documentclass[11pt,a4paper]{article}
\pdfoutput=1
\usepackage{jheppub}
\usepackage[utf8]{inputenc} 
\usepackage{multirow}    
\usepackage{xcolor}
\usepackage{graphicx}
\usepackage{amsmath}
\usepackage{amssymb}
\usepackage{braket} 
\usepackage{hyperref}
\usepackage{frontespizio}
\usepackage{comment}  
\usepackage{epsfig}
\usepackage{float}
\usepackage{bbold}
\usepackage[title]{appendix}
\usepackage{multirow} 
\usepackage{tikz}
\usetikzlibrary{decorations.pathmorphing}
\usepackage{varwidth}
\usepackage{tikz}
\usetikzlibrary{backgrounds,patterns,calc}
\usepackage{xparse}  
\usepackage{subfigure}  
\usepackage{bbm}   
\usepackage{mathabx}
\usepackage{natbib}

\usepackage{jheppub} % for details on the use of the package, please
\title{Drinfeld Centers from Magnetic Quivers}

\author[]{Veronica Pasquarella}%{\ and Fernando Quevedo}}
\affiliation{Department of Applied Mathematics and Theoretical Physics (DAMTP)}
\affiliation{University of Cambridge, \\ Wilberforce Road, CB3 0WA, Cambridge, UK}  

\emailAdd{vp360@damtp.cam.ac.uk}%, fq201@damtp.cam.ac.uk}

\abstract{The present work shows that magnetic quivers encode the necessary information for determining the Drinfeld center in the symmetry topological field theory constructions (SymTFT) associated to a given absolute theory. The crucial argument resides in their common aim of generalising homological mirror symmetry.}


\makeatletter
\gdef\@fpheader{}
\makeatother
\begin{document} 
\maketitle
%\flushbottom 

\section{Introduction}

This is the first of two papers by the same author addressing the formulation of mirror symmetry from the perspective of geometric representation theory. Mostly inspired by \cite{Teleman:2014jaa} and further upcoming work by Teleman, \cite{CT}, the present article shows how constructions as the ones addressed in \cite{Pasquarella:2023deo}, involving multiple gaugings-by-condensation, can still be assigned a fiber functor, as in the Freed-Moore-Teleman symmetry TFT (SymTFT) setup, \cite{Freed:2022qnc}; however, such fiber functor should not be associated to either of the absolute theories separated by the non-invertible defect, but, rather to a 3D theory whose partition function correctly accounts for the symplectic projection leading to either of the two absolute theories.

Our approach consists in a combination of the following:

\begin{enumerate} 

\item Defining a fiber functor for different absolute theories connected by non-invertible defects, \cite{Pasquarella:2023deo}. 

\item Relating mirror symmetry with the identification of a Drinfeld center, \cite{Teleman:2014jaa,CT}. 

\item Recent advancements in the understanding of the Higgs branch (HB) structure of higher-dimensional quiver gauge theories with 8 supercharges by means of Coulomb branches (CBs) of magnetic quivers (MQs) associated to 3D ${\cal N}=4$ gauge theories, \cite{Cabrera:2019izd,Bourget:2019aer,Bourget:2021siw,Bourget:2023cgs,Bourget:2019rtl,Ferlito:2016grh}. 

\end{enumerate}

In this first work, we mostly outline how the topics listed above combine together, and in the second, \cite{VP1}, we will provide a more detailed mathematical correspondence with the work of \cite{Teleman:2014jaa}, highlighting how his constructions can effectively be generalised to higher dimensions and be naturally embedded in the setup of \cite{Freed:2022qnc} precisely thanks to the understanding in terms of magnetic quivers of 3D ${\cal N}=4$ theories addressed here.  

One could see our proposal as further supporting the idea that higher-categorical symmetries probe representation theory structures.

The present work is structured as follows: in section \ref{sec:DCFMQS} we review the correspondence between geometric and algebraic resolutions of framed Nakajima quiver varieties, \cite{DAlesio:2021hlp,Braverman:2016pwk,Braverman:2016wma}, highlighting it as an interesting example of homological mirror symmetry. In particular, we emphasise the property the moment map and higher homologies need to satisfy to ensure agreement in between the calculation of the two Hilbert series. We conclude the section with a brief overview of Hasse diagram constructions via magnetic quivers for quiver gauge theories with 8 supercharges, pointing out an interesting 2-categorical structure arising when dealing with complete intersections. 

In section \ref{sec:3} we explain how gauging-by-condensation can be related to the poset ordering leading to the construction of Hasse diagrams, thanks to the unifying role of the moment map. We then explain how the identification of such moment map ensures the quiver gauge theory enjoys a generalised notion of homological mirror symmetry, with the latter corresponding to the presence of a Drinfeld center and a corresponding fiber functor for a 2-categorical structure, related to Rozansky-Witten theory, \cite{Rozansky:1996bq}. We conclude highlighting connections between the topics outlined in the present work and those of \cite{Teleman:2014jaa, CT}, thereby opening the scene to the more mathematical treatment to which \cite{VP1} is devoted.

\section{Magnetic Quivers: a unifying framework}   \label{sec:DCFMQS}

This first section is meant to provide a brief overview of some key elements we will be using throughout our treatment, emphasising the ones that are needed for building the connection with the work of \cite{Teleman:2014jaa, CT}\footnote{Throughout the entire treatment, we will be assuming the basic knowledge of higher categories and ADE quivers. We refer the reader to the extended literature on both topics for detailed definitions and examples. Specific additional tools will be explained in due course when needed.}, which will be the core focus of section \ref{sec:3}. The first part of the present article is structured as follows: 

\begin{itemize}  

\item  At first we review the correspondence between geometric and algebraic resolutions of framed Nakajima quiver varieties, \cite{DAlesio:2021hlp,Braverman:2016pwk,Braverman:2016wma}, highlighting it as an interesting example of homological mirror symmetry\footnote{The main explanation for this statement will become manifest in section \ref{sec:3}.}. In particular, we emphasise the property the moment map and higher homologies need to satisfy to ensure agreement in between the two resolutions. Consequently, this is mapped to an equivalence in between the Hilbert series resulting from summing over characters.

\item We then turn to a brief overview of Higgs and Coulomb branches\footnote{Simply denoted by HB and CB for convenience.} of quiver gauge theories as algebraic varieties, pointing out the role of magnetic quivers (MQs) for describing the HB of a quiver gauge theory with 8 supercharges and arbitrary flavour, \cite{Cabrera:2019izd,Bourget:2019aer,Bourget:2021siw,Bourget:2023cgs,Bourget:2019rtl,Ferlito:2016grh}. In doing so, we highlight the importance of the role played by, both, the Hilbert series (HS) and highest weight generating function (HWG) in identifying different intersecting cones in the HB Hasse diagram, \cite{Cabrera:2019izd,Bourget:2019aer,Bourget:2021siw,Bourget:2023cgs,Bourget:2019rtl,Ferlito:2016grh}, suggesting a 2-categorical structure when dealing with complete intersections.  

\end{itemize}   

As we shall see, the underlying motivation for building the connections explained in section \ref{sec:3} is the importance of identifying a unifying framework for realising mirror symmetry and its generalisations.

\subsection{Framed Nakajima quiver varieties}  \label{sec:2.1}

In this first subsection, we briefly overview the features of framed Nakajima quiver varieties, \cite{DAlesio:2021hlp}, and their geometric and algebraic resolutions.

\subsection*{\texorpdfstring{$\bullet$}{}\ \ \ Geometric (Nakajima) resolution }

\begin{figure} [ht!]  
\begin{center}
\includegraphics[scale=1]{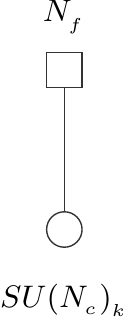}  
\caption{\small An example of a framed Nakajima quiver variety. This is a prototypical example of the theories we will be addressing in the present work, namely for quiver gauge theories with 8 supercharges. The unframed quiver would be the gauge node without matter. Here we have set $G_{_v}\equiv SU(N)$. However, in principle, other choices could have been taken.}  
\label{fig:FNQV}  
\end{center}  
\end{figure}

Quiver varieties are varieties of quiver representations of a quiver: one fixes a vector space at each vertex, then considers the linear space of representations associated to each arrow of the quiver linear map. A framed version of this was initially introduced by Kronheimer and Nakajima, amounting to doubling the set of vertices, and drawing a new arrow from each new vertex to its corresponding old one. An example of the resulting quiver (which is of the prototypical type of interest for us) is depicted\footnote{The nomenclature featuring in the framed quiver will be explained in section \ref{sec:2.2}, but we emphasise that this is standard notation in the quiver literature.} in figure \ref{fig:FNQV}.  

Framed representations also appear naturally in ADHM quiver constructions of self-dual or anti-self-dual YM on $S^{^4}$, \cite{Atiyah:1978ri}. This is particularly interesting from the point of view of representation theory of Lie algebras because dimension vectors of the framed vertices appear as highest weights of the representations. There are different notions of framing. According to Nakajima's version for quiver varieties, the framed quiver is doubled, meaning each arrow gets doubled by an arrow that goes in the opposite direction. The linear space of representations becomes a linear cotangent bundle

\begin{equation}
    M({\cal Q},v,w)\ \overset{def.}{=}\ T^{^*} L({\cal Q}^{^{\text{fr}}},v,w),   
\end{equation}
where ${\cal Q}, {\cal Q}^{^{fr}},v,w$ respectively denote the original quiver, the framed quiver, the number of original vertices, and the number of framed vertices.
The gauge group is a general linear group on the original vertices  $G\equiv G_{_v}$, and there is a moment map 

\begin{equation}   
\boxed{\ \ \ \ \ \mu:\ M({\cal Q},v,w)\ \rightarrow\ \mathfrak{g}^{^*} \color{white}\bigg] \ \ },   
\end{equation}  
with $\mathfrak{g}$ denoting the Lie algebra of the theory. Nakajima quiver varieties are Hamiltonian reductions of the following action 

\begin{equation} 
G\  \circlearrowright\ M({\cal Q},v,w),   
\end{equation}
and come into two types:  

\begin{enumerate} 

\item the \emph{affine} Nakajima variety, defined as the partial character variety

\begin{equation}   
\mathfrak{M}^{^0}({\cal Q},v,w)\ \equiv\ \mu^{^{-1}}(0)//G,  
\label{eq:first}
\end{equation} 

\item \emph{quasi-projective}, also known as the GIT quotient, \cite{GIT},

\begin{equation}   
\mathfrak{M}^{^{\chi}}({\cal Q},v,w)\ \equiv\ \mu^{^{-1}}(0)//_{_{\chi}}G.   
\label{eq:second}
\end{equation} 

\end{enumerate}

The gauge group by which we take the quotient is   

\begin{equation}  
G\ \equiv\ G_{_v}\ \overset{def.}{=}\ \prod_{a}\ GL_{_{v_{a}}} (\mathbb{C})\ \subset G_{_{v}}\ \times G_{_{w}}   
\end{equation}

For any choice of the nontrivial character    

\begin{equation}  
\chi:\ G\ \rightarrow\ \mathbb{C}^{^{\times}} 
\end{equation}   
there is a proper Poisson morphism 

\begin{equation} 
p: \mathfrak{M}^{^{\chi}}({\cal Q},v,w)\ \rightarrow\ \mathfrak{M}^{^0}({\cal Q},v,w),  
\end{equation}  
which is as a symplectic resolution of the singularities of $\mathfrak{M}^{^0}$.

In both cases, \eqref{eq:first} and \eqref{eq:second}, $\mu^{^{-1}}(0)$ denotes the fiber of zero through the moment map. The latter can be used as a representation scheme for the path algebra, ${\cal A}$, modulo the ideal, ${\cal I}_{_{\mu}}$,

\begin{equation}   
{\cal A}\ \overset{def.}{=}\  \mathbb{C}\overline{{\cal Q}^{^{fr}}}/{\cal I}_{_{\mu}},   
\end{equation}
of the framed doubled quiver 

\begin{equation}   
\mu^{^{-1}}(0)\ \equiv\ \text{Rep}_{_{\mathbb{C}^{^v}\otimes\mathbb{C}^{^w}}}({\cal A})\ \overset{def.}{=}\  \text{Rep}_{_{v,w}}({\cal A}),   
\end{equation} 
where we made use of the following shorthand notation

\begin{equation}   
\mathbb{C}^{^v}\ \overset{def.}{=}\ \bigoplus_{a}\ \mathbb{C}^{^{v_{_a}}}\ \ \ , \ \ \ \mathbb{C}^{^w}\ \overset{def.}{=}\ \bigoplus_{a}\ \mathbb{C}^{^{w_{_a}}},   
\end{equation}

For the purpose of our work, the main result of \cite{DAlesio:2021hlp} is being able to relate such varieties to derived representation schemes, with the latter being obtained by means of an alternative resolution of the affine Nakajima quiver variety. The underlying reason for the importance of this resides with it being a sample realisation of homological mirror symmetry\footnote{We will be explaining this in due course.}. 

Prior to explaining the alternative resolution leading to derived representation schemes, we wish to emphasise that the correspondence that we are looking for, summarised in figure \ref{fig:GIT}, \cite{DAlesio:2021hlp}, emerges from comparing invariants, and the required conditions for them to match at the two endpoints of the red arrow. From the symplectic resolution point of view, this requires flatness of the moment map, as explained in \cite{DAlesio:2021hlp}, following \cite{mommap}. We refer the interested reader to such references for more detailed explanation regarding the definition and properties of flat moment maps. For our purposes, the crucial point is that when the invariants in the symplectic and derived scheme description match, we are dealing with a complete intersection\footnote{The meaning of the latter will be explained in section \ref{sec:2.2}.}.

\subsection*{\texorpdfstring{$\bullet$}{}\ \ \ Algebraic resolution}

The technique of algebraic resolution was first put forward by \cite{Berest}, and consists in resolving the singularities of the representation schemes by introducing homological algebra. 

Prior to explaining derived representation schemes, let us first briefly overview the notion of a differential graded (dg)-schemes. A dg-scheme is a pair $X\overset{def.}{=}\left(X_{_o},{\cal O}_{_{X,\bullet}}\right)$, with $X_{_o}\equiv M({\cal Q},v,w)$ denoting the vector space of linear representations, and ${\cal O}_{_{X,\bullet}}$ a sheaf of dg-algebras such that their zeroth homology reads as follows 

\begin{equation}   
\pi_{_0}(X)\ \overset{def.}{\equiv}\ \text{Spec}\left(H_{_0}\left({\cal O}_{_{X,\bullet}}\right)\right)\ \equiv\ \mu^{^{-1}}(0). 
\end{equation}

The derived representation scheme of the relative algebra\footnote{Associated to the choice of a path in the framed quiver.}, ${\cal A}$, in a vector space $V$, is the object, \cite{DAlesio:2021hlp},  

\begin{equation} 
\text{DRep}_{_V}({\cal A})\ \in\ \text{Ho}\left(\text{DGAff}_{_k}\right), 
\end{equation}
obtained applying the following composition of functors  

\begin{equation} 
\text{DRep}_{_V}(-):\ \text{Alg}_{_S}\ \longrightarrow\ \text{Ho}(\text{DGA}_{_S}^{^+})\ \xrightarrow{L(-)_{_V}}\ \text{Ho}(\text{CDA}_{_k}^{^+})\ \xrightarrow{\text{RSpec}}\ \text{Ho}(\text{DGAff}_{_k}),
\label{eq:totalmap}    
\end{equation}
where   

\begin{equation}   
L(-):\ \text{Ho}\left(\text{DGA}_{_S}\right)\ \rightarrow\ \text{Ho}\left(\text{CDGA}_{_S}\right)  
\end{equation}
is the derived representation functor acting on the homotopy category, Ho, whose homology is the representation homology of a differential graded algebra, ${\cal A}$, 

\begin{equation}    
H_{_{\bullet}}({\cal A},V)\ \overset{def.}{=}\ H_{_{\bullet}}\left(L({\cal A})_{_V}\right).  
\end{equation}

On the other hand, RSpec in \eqref{eq:totalmap} denotes the derived spectrum, defining an equivalence on the homotopy categories 

\begin{equation}
\includegraphics[scale=1.1]{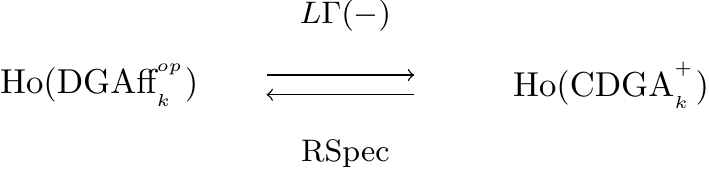}. 
\end{equation}

Altogether, \eqref{eq:totalmap} leads to 
\begin{equation} 
\text{DRep}_{_V}({\cal A})\ =\ \text{RSpec}\left(L({\cal A}_{_V})\right) =\ \text{Rep}_{_V}\left({\cal A}_{_{cof}}\right),    
\label{eq:DRS}   
\end{equation}
with 

\begin{equation} 
{\cal A}_{_{cof}}\ \overset{\sim}{\twoheadrightarrow}\ {\cal A}
\end{equation}
a cofibrant replacement. Different choices of cofibrant replacements lead to different models of \eqref{eq:DRS}. For the specific case of framed Nakajima quiver varieties, the derived representation scheme reads, \cite{DAlesio:2021hlp},  

\begin{equation} 
\text{DRep}_{_{v,w}}({\cal A})\ =\ \text{Spec}\left(L({\cal A}_{_{v,w}})\right) \ \in\ H_{_0}\left(\text{DGAff}_{_{\mathbb{C}}}\right), 
\label{eq:DRSDA}   
\end{equation}
with representation homology 

\begin{equation} 
H_{_{\bullet}}({\cal A},v,w)\ \equiv\ H_{_{\bullet}}\left(L({\cal A})_{_{v,w}}\right)\ \in\ \text{CDGA}_{_{\mathbb{C}}}^{^+}, 
\label{eq:DRSDA1}   
\end{equation}
a graded commutative algebra. Denoting by $\{U_{_{\lambda}}\}$ the irreducible representations of $G$, \eqref{eq:DRSDA1} explicitly reads as follows 

\begin{equation} 
H_{_{\bullet}}({\cal A},v,w)\ \overset{def.}{=}\ \underset{\lambda}{\bigoplus}\ \text{Hom}_{_G}\left(U_{_{\lambda}}, H_{_{\bullet}}({\cal A},v,w)\right)\otimes U_{_{\lambda}}, 
\label{eq:DRSDA2}   
\end{equation}  
with $\text{Hom}_{_G}\left(U_{_{\lambda}}, H_{_{\bullet}}({\cal A},v,w)\right)$ denoting its isotypical components, namely modules over

\begin{equation} 
H_{_0}({\cal A},v,w)^{^G}\ \equiv\ {\cal O}\left(\mu^{^{-1}}(0)\right)^{^G}.
\label{eq:DRSDA3}   
\end{equation}

For each irreducible representation of $G_{_v}$, the Euler character is defined as follows  

\begin{equation}  
\begin{aligned}
\chi_{_T}^{^{\lambda}}({\cal A},v,w)\ &\overset{def.}{=}\ \overset{\infty}{\underset{i=0}{\sum}}\ (-1)^{^i}\left[\text{Hom}_{_G}\left(U_{_{\lambda}}, H_{_i}({\cal A},v,w)\right)\right]\\
&=\overset{\infty}{\underset{i=0}{\sum}}\ (-1)^{^i}\left[H_{_i}(L({\cal A}))_{_{\lambda,v,w}}^{^{G_{_v}}}\right]\ \ \ \in\ \ K_{_T}\left(\mathfrak{M}^{^0}\right),  
\end{aligned}
\end{equation}
where $T\overset{def.}{=}T_{_w}\times T_{_l}$ accounts for the framing, and 

\begin{equation}  
H_{_i}(\pi):\ H_{_i}\left({\cal A}_{_{cof}}\right)\ \rightarrow\ H_{_i}({\cal A})  
\end{equation}  
is an isomorphism such that ${\cal A}_{_{cof}}$ has no higher homologies, namely, 

\begin{equation}  
\begin{cases} 
&H_{_0}(\pi):\ H_{_i}\left({\cal A}_{_{cof}}\right)\ \xrightarrow{\sim}\ {\cal A}\\    
&\\    
&H_{_i}\left({\cal A}_{_{cof}}\right)\ \equiv\ 0, \ \forall i\ge1.\\
\end{cases}
\end{equation} 
and 

\begin{equation} 
\pi:\ {\cal A}_{_{cof}}\ \longrightarrow\ {\cal A}   
\end{equation} 
defines an acyclic fibration in DGA$_{_{\mathbb{C}}}^{^+}$.

\subsection*{\texorpdfstring{$\bullet$}{}\ \ \ Hilbert series}

One of the key results of \cite{DAlesio:2021hlp} is that agreement in between invariants calculated from the Nakajima and derived representation schemes requires flatness of the moment map on the symplectic side, and vanishing higher homologies in the derived case. Importantly for us, such agreement involves integrated characters, also known as the Hilbert series (HS). The latter is a map which reads

\begin{equation} 
\text{HS}_{_T}:
\begin{cases} 
&K_{_T}\left(\mathfrak{M}^{^0}\right)\ \longrightarrow\ K_{_T}(\text{pt}),\  \text{if $\mathfrak{M}^{^0}$ is compact}\\ 
&\\ 
&K_{_T}\left(\mathfrak{M}^{^0}\right)\ \longrightarrow\ \text{Frac}(T),\  \text{otherwise}.\\ 
\end{cases}
\end{equation}   

From this, the resulting Weyl integral formula reads 

\begin{equation} 
\text{HS}_{_T}\left(\chi_{_T}^{^G}({\cal A},v,w)\right)\ \equiv\ \frac{1}{|G|}\ \int_{_G}dg\ \text{HS}_{_{G\times T}}\left(\chi_{_T}^{^G}({\cal A},v,w)\right).
\label{eq:hs}  
\end{equation}

Upon choosing the variables in the maximal torus of the gauge group, $x\ \in\ T_{_v}\ \subset\ G$, and equivariant variables $t\equiv\ (a,l)\ \in\ T\ \equiv\ T_{_w}\times T_{_l}$, \eqref{eq:hs} can be re-expressed as follows

\begin{equation} 
\text{HS}_{_T}\left(\chi_{_T}^{^G}({\cal A},v,w)\right)\ \equiv\ \frac{1}{|W|}\ \int_{_{T_{_v}}}dx\ WF(x)\ \frac{\ \underset{i}{\prod}\ \left(1-l_{_1}l_{_2}b_{_i}\right)\ }{\underset{j}{\prod}\ \left(1-a_{_j}\right)}, 
\label{eq:hs1}  
\end{equation}   
with $b_{_i}, a_{_i}$ respectively denoting the weights of $M({\cal Q},v,w)^{^*}$ and $\mathfrak g$.

\subsection*{\texorpdfstring{$\bullet$}{}\ \ \ Homological mirror symmetry}

Mirror symmetry constitutes an active area of research within the theoretical physics community, specially motivated by string theory. Seeking for a mathematical formulation that could explain the origin of such symmetry has been, and still is, an active area of research within the mathematical community, where the correspondence is also referred to as homological mirror symmetry, \cite{Hori:2000kt,Klemm:1996bj,Kontsevich:1994dn}. The latter consists in a proposed agreement between two categories, namely:

\begin{itemize}  

\item  Fukaya's $A_{_{\infty}}$-category, ${\cal F}(X)$, on the symplectic side,

\item and the derived category with Yoneda structure, $D^{^b}\mathfrak{Coh}(X^{^{\text{V}}})$, on the complex side. 

\end{itemize} 

For completeness, we now turn to explain this terminology.
A symplectic manifold, $X$, is a smooth manifold of even dimension equipped with a non-degenerate symplectic form $\omega\ \in\ \Omega^{^2}_{_{cl}}(X)$. A Fukaya category of a symplectic manifold, $X$, is an $A_{_{\infty}}$-category with Lagrangian submanifolds $L\ \subset\ X$ as objects, whose intersections define the Hom-space. 
An $A_{_{\infty}}$-category is a category with associativity condition ($A)$ relaxed without bound on degrees of homotopies ($\infty$). They are linear categories, i.e. their Hom-objects are chain complexes. A Lagrangian submanifold of a symplectic manifold, $L\ \subset\ X$, is a submanifold which is a maximal isotropic submanifold on which $\omega=0$. They constitute the leaves of real polarisations, and are, therefore, crucial elements of symplectic geometry.

On the other hand, a Yoneda structure is a pair $(\mathbb{B},{\cal P})$ on a 2-category ${\cal K}$, with $\mathbb{B}$ an admissible class of 2-cells and ${\cal P}$ a presheaf construction for $\mathbb{B}$, assigning to every object $B\in\ \mathbb{B}$ an object of presheaves ${\cal P}B\ \in\ {\cal K}$, and a Yoneda morphism $B\rightarrow{\cal P}B$.

\begin{figure}[ht!]  
\begin{center} 
\includegraphics[scale=1]{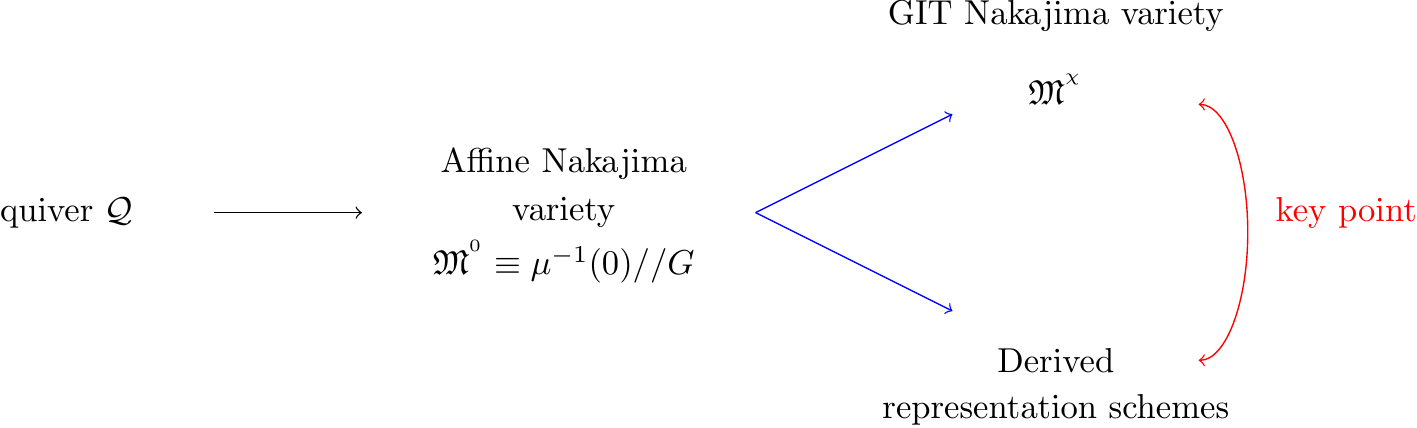} 
\caption{\small The upper and lower blue arrows correspond to geometric and algebraic resolutions of ${\mathfrak{M}^{0}}$, respectively. The red arrow corresponds to a realisation of homological mirror symmetry.}  
\label{fig:GIT}  
\end{center}    
\end{figure}

Within the context of framed Nakajima quiver varieties, such correspondence is mapped to an equivalence between the category of $\mathbb{C}$-linear representations of a quiver, ${\cal Q}$, and the category of left $\mathbb{C}{\cal Q}$-modules, as summarised in figure \ref{fig:GIT}.

    \subsection*{\texorpdfstring{$\bullet$}{}\ \ \ Moment map and higher homologies}

In general, homologies of derived representation schemes can be highly nontrivial. However, in this particular case, one can identify a necessary and sufficient condition for the vanishing of the higher homologies based on the flatness of $\mu$, \cite{DAlesio:2021hlp}. In particular, in such reference, it was shown that the derived representation scheme DRep$_{_{v,w}}({\cal A})$ has vanishing higher homologies if and only if $\mu^{^{-1}}(0)\ \subset\ M({\cal Q},v,w)$ is a complete intersection, which happens only if the moment map is flat, \cite{mommap}. As we shall see, the requirement of the algebraic variety to be a complete intersection is crucial for the purpose of our treatment. In particular, it ensures the emergence of a 2-categorical structure, whose importance will be the core topic of section \ref{sec:3}. 

Prior to that, we devote the following subsection to explain the role played by magnetic quivers for describing generalisations of homological mirror symmetry in quiver gauge theories with 8 supercharges.

\subsection{Intersecting cones} \label{sec:2.2}  

From the more mathematically-inclined overview of section \ref{sec:2.1}, we know how to obtain a configuration admitting homological mirror symmetry, and the conditions that must be satisfied for the Hilbert series to match on the GIT and derived representation scheme side. For the purpose of what we will be addressing in section \ref{sec:3}, we will now overview a setup where homological mirror symmetry can be described in an interestingly generalised and controlled manner, suggesting this as a probe for investigating its underlying mathematical structure.

\begin{figure}[ht!]     
\begin{center}
\includegraphics[scale=1]{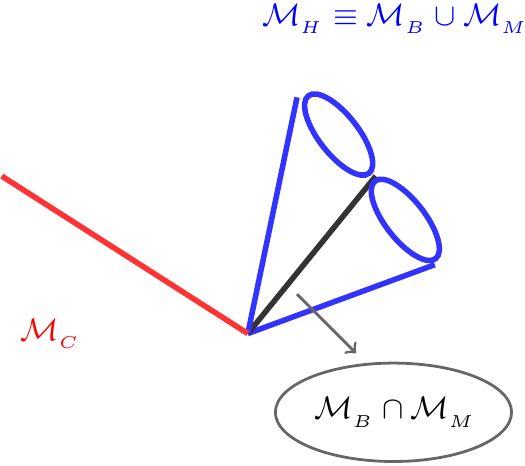} 
\caption{\small Moduli space of a supersymmetric theory, with a Coulomb branch (CB), ${\cal M}_{_C}$, and a Higgs branch (HB), ${\cal M}_{_H}\equiv{\cal M}_{_B}\cup{\cal M}_{_M}$, with its nontrivially intersecting mesonic and baryonic branches, denoted by ${\cal M}_{_B}$ and ${\cal M}_{_M}$, respectively.}  
\end{center} 
\end{figure} 

%Key idea: stratification of the null cone via the moment map [Ness, Mumford, 1984]. The basic problem of geometric invariant theory is the construction of orbit spaces.

\subsection*{Higgs and Coulomb branches as algebraic varieties}

The main focus of this section are quiver gauge theories with 8 supercharges, classical flavour and gauge groups, whose vacuum moduli space is related to the nilpotent orbits, ${\cal O}_{_{\rho}}$, of classical Lie algebras, $\mathfrak{g}$. For a given gauge group $G$, the nilpotent orbits of an algebra $\mathfrak{g}$ are defined by equivalence classes of nilpotence conditions on representation matrices, \cite{Hanany:2019tji}. From the theoretical physics point of view, the nilpotent conditions describe the way in which the scalar fields in the $F$-term equations\footnote{Namely the derivatives of the superpotential.} vanish at the supersymmetric vacuum, and can be specified by a quiver gauge theory. 

According to the Jacobson Morozov Theorem, these nilpotent orbits are 
in one to one correspondence with equivalence classes of embeddings of $\mathfrak{su}(2)$ into $\mathfrak{g}$. Each such embedding 

\begin{equation}   
\rho:\mathfrak{su}(2)\ \rightarrow\ \mathfrak{g}  
\end{equation}   
is a homomorphism, labelled by, either, the partition of the representation of $G$, or by a characteristic, using Dynking labels to specify the mapping of the roots and weights of $\mathfrak{g}$ onto $\mathfrak{su}(2)$.

A \emph{Slodowy slice}, $S_{_{\rho}}\overset{def.}{\equiv}{\cal O}_{_{\rho}}^{^{\perp}}$, is a space transverse to a nilpotent orbit, and therefore commuting with it, while living within the adjoint orbit of the ambient group $G$. These transverse spaces can be further restricted to their intersections with the closure of any enclosing nilpotent orbit $\overline{\cal O}_{_{\sigma}}$, thereby leading to spaces labelled by pairs of nilpotent orbits, whose elements are \emph{Slodowy interesections}\footnote{The connection between the 3D boundary conditions on type-II brane systems in 4D ${\cal N}=4$  CFTs and Slodowy intersections was highlighted in \cite{Gaiotto:2008ak}.} 

\begin{equation}    
S_{_{{\sigma,\rho}}}\ \overset{def.}{\equiv}\ \overline{\cal O}_{_{\sigma}}\ \cap\ S_{_{\rho}},  
\label{eq:SlInt}    
\end{equation}    
encoding, as particular examples, nilpotent orbits, Slodowy slices and Kraft-Procesi transitions, \cite{KP,KP1}, with the latter being defined by intersections, ${\cal S}_{_{\rho^{\prime},\rho}}$, between pairs of orbits $(\rho^{\prime},\rho)$ that are adjacent in a Hasse diagram\footnote{This will become clearer when introducing the notion of quiver subtraction in the procedure for recovering the Hilbert series associated to a given electric quiver.}. \eqref{eq:SlInt} are the algebraic varieties of which a Hilbert series can be calculated\footnote{Hilbert series for Slodowy intersections can also be constructed by means of purely group theoretic methods, making use of localisation formulae related to the Hall Littlewood polynomials.}, as we will explain later on. Prior to turning to that, we outline some preliminary definitions that will be recursively used throughout the remainder of this work.

The \emph{nilcone} is defined as the closure of the maximal nilpotent orbit,

\begin{equation}    
\overline{\cal O}_{_{max}}\ \overset{def.}{\equiv}\  {\cal N},      
\end{equation} 
with dimension

\begin{equation}    
\left|{\cal N}\right|\ \equiv\ \left|\mathfrak{g}\right|-\text{rank}\left[\mathfrak{g}\right].       
\end{equation}

Nilpotent orbits can be arranged as a \emph{Hasse diagram} according to the inclusion relations of their closures:  

\begin{equation}    
{\cal N}\ \equiv\ \overline{\cal O}_{_{max}}\ \supset\  \overline{\cal O}_{_{sub-reg}}\ \supset\ ...\ \overline{\cal O}_{_{min}}\ \supset\ \overline{\cal O}_{_{trivial}}\ \equiv\ \{0\}.    
\end{equation}  

From the definition of the Slodowy intersection, it follows that

\begin{equation}    
S_{_{{{\cal N},trivial}}}\ \equiv\ \overline{\cal O}_{_{max}}   \equiv\ {\cal N}
\end{equation}

By means of such terminology, the definition of the \emph{Higgs} and \emph{Coulomb branches}\footnote{As previously anticipated, we will simply denote them as HB and CB, respectively.} can be expressed as follows
\begin{equation}    
\overline{\cal O}_{_{max}}\   \overset{def}{\equiv}\ \text{HB}\left[{\cal M}_{_A}(\rho,0)\right] 
\ \ \ , \ \ \      
S_{_{{{\cal N},\rho^{^T}}}}\   \overset{def}{\equiv}\ \text{CB}\left[{\cal M}_{_A}(\rho,0)\right] 
\ \ \ , \ \ \     
{\cal M}_{_A}(\rho,0)\   \overset{def}{\equiv}\ {\cal L}_{_{A}}\left(\rho^{^T}\right)
\end{equation}  
with ${\cal M}_{_A}(\rho,0)$ denoting a single-flavoured linear quiver.

\subsection*{Magnetic quivers and Hasse diagrams}  

Having said this, we now briefly overview the prescription for constructing magnetic quivers and Hasse diagrams from quiver subtraction, \cite{Cabrera:2018ann}\footnote{Such terminology is explained in due course. We refer to the extensive literature, especially the one in the references, for more detailed explanations as well as several examples explicitly analysed.}. The procedure can be summarised by the following steps:  

\begin{enumerate}  

\item Start from a certain quiver gauge theory with 8 supercharges. Focussing on SQCD theories, namely a 3-parameter family of theories labelled by $(N_{_c}, N_{_f}, k)$, denoting number of colours, flavours, and Chern-Simons level, respectively. As a framed Nakajima quiver variety, we assign to it an electric quiver of the kind depicted in figure \ref{fig:FNQV}. Its HB is an algebraic variety associated to a 5-brane web configuration. 

\item Identify the number of maximal decompositions of the 5-brane web. 

\item To each maximal decomposition associate the magnetic quiver (MQ) of a 3D ${\cal N}=4$ gauge theory.  

\item The HB of the SQCD theory is preserved under dimensional reduction, and is equivalent to the union of the CBs of the MQs identified in step 3.    

\item Construct the Hasse diagram associated to the HB by implementing quiver subtraction\footnote{As previously mentioned, we are assuming familiarity with the notion of ADE quivers. We refer to the extended literature on the topic if needed.} on the MQs.  

\end{enumerate}   

Most importantly, this procedure generalises the following relation:  

\begin{equation}  
\boxed{\ \ \ \ \text{HB}\left(T\right)\ \equiv\ \text{CB}\left(T^{^{\text{V}}}\right)\color{white}\bigg]\ \ }, 
\label{eq:mirrordual}  
\end{equation} 
with $T$ a 3D ${\cal N}=4$ quiver gauge theory, and $T^{^{\text{V}}}$ its mirror dual. \eqref{eq:mirrordual} can also be re-expressed in terms of the electric and magnetic quivers 

\begin{equation}  
\boxed{\ \ \ \ \text{HB}\left(\text{EQ}\right)\ \equiv\ \text{CB}\left(\text{MQ}\right)\color{white}\bigg]\ \ }. 
\label{eq:mirrordualquivers}  
\end{equation} 

The generalised expressions for \eqref{eq:mirrordual} and \eqref{eq:mirrordualquivers} would therefore be  
\begin{equation}  
\boxed{\ \ \ \text{HB}\left(T\right)\ \equiv\ \underset{i}{\bigcup}\ \text{CB}\left(\tilde T_{_i}^{^{\text{V}}}\right) \ \ \ ,\ \ \   
\text{HB}\left(\text{EQ}\right)\ \equiv\ \underset{i}{\bigcup}\ \text{CB}\left(\text{MQ}_{_i}\right)\color{white}\bigg]
\ \ \ }   .    
\label{eq:genms}         
\end{equation}

The advantage of the procedure put forward by \cite{Cabrera:2019izd,Bourget:2019aer,Bourget:2021siw,Bourget:2023cgs,Bourget:2019rtl,Ferlito:2016grh} is that, when $T$ does not admit a unique $T^{^{\text{V}}}$, its \emph{generalised} mirror symmetric theory can still be recovered by specifying the MQs, or, equivalently, the generators of the cones featuring in the Hasse diagram, together with their intersections. For concreteness, we briefly overview some key examples of the procedure summarised above, encompassing the main features we will be needing to consider in section \ref{sec:3}.

\subsection*{Rank-1 and rank-2 examples}

The examples reproduced in this subsection are for 5D ${\cal N}=1$ SQCD, with varying parameters $(N_{_c},N_{_f},k)$. Notice that, for $SU(2)$, $k=0$, reason why it is conventionally omitted in the quiver representation. However, for $N_{_c}>2$ it might be $k\neq0$, reason why in such case we specify this in the corresponding quiver.

For each case we draw brane intersections (NS5s-branes and (p,q)-branes are always present). When adding flavour, D5-branes also contribute to the intersection. On the right of each brane intersection, we show the corresponding Hasse diagram constructed by implementing quiver subtraction on the magnetic quiver(s) associated to the brane decomposition(s). Conventionally, the node at the bottom of the Hasse diagram denotes the original magnetic quiver, prior to implementing any quiver subtraction. At each step of the iteration, we draw a vertical line going upwards, ending at another node, with the latter denoting a reduced magnetic quiver. When a bifurcation takes place, it signals that there are multiple possible quiver subtractions that could be performed. This might lead to the emergence of multiple cones in the Hasse diagram. Conventionally, the nodes and lines featuring in a Hasse diagram ar referred to as symplectic leaves and symplectic slices, respectively. We will show this by going through some low-rank examples.

\subsection*{\texorpdfstring{$\bullet \ \ \text{5D}\ \ SU(N_{_c})_{_0}, N_{_f}=0$}{}}

\medskip 

\medskip 

This first example has no flavour degrees of freedom. Hence, the 5-brane web only consists of intersecting NS5- and $(p,q)$-branes, as shown on the LHS of figure \ref{fig:2p}. Its unique decomposition means there is a unique magnetic quiver associated to this theory, leading to equation \eqref{eq:firstcase}\footnote{Here we are using the notation of \cite{Cabrera:2019izd} where the gauge nodes of the MQs are simply labelled by the ranks of unitary gauge nodes.}.

\begin{equation} 
\includegraphics[scale=0.6]{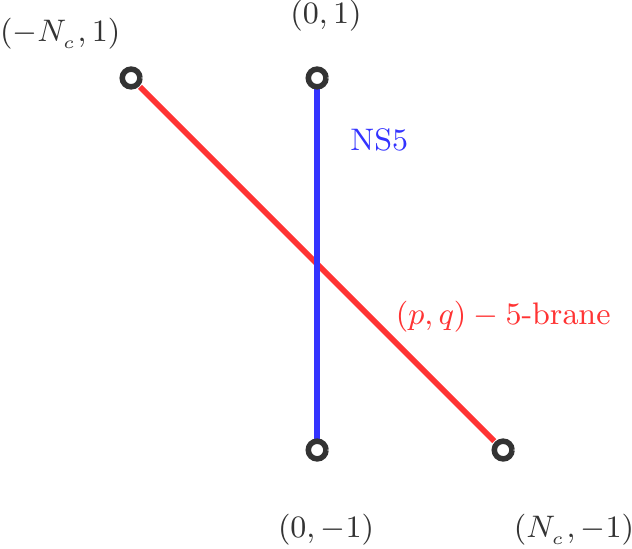} 
\qquad\qquad\qquad\qquad\qquad  
 \includegraphics[scale=0.9]{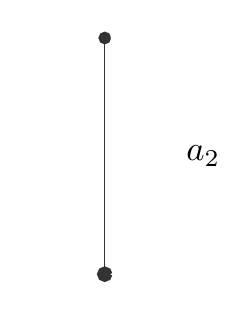} 
\nonumber
\end{equation}

\begin{figure}[ht!]    
\begin{minipage}[c]{1\textwidth}
\caption{\footnotesize In absence of flavour degrees of freedom, the 5-braneweb decomposition is unique (LHS) leading to a HB Hasse diagram characterised by a unique cone (RHS). }     \label{fig:2p}   
\end{minipage}    
\end{figure}
\medskip 

\medskip 

\begin{equation}  
\text{HB}_{_{\infty}}^ {\ \text{5D}}\ \left(\ \raisebox{-30pt}{\includegraphics[scale=0.75]{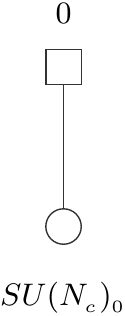}}\ \right) \ \equiv\ \mathbb{C^{^3}}/\mathbb{Z}_{_N{_{_c}}}\ \equiv \ \text{CB}^{\ \text{3D}}\ \left(\ \raisebox{-15pt}{\includegraphics[ scale=0.75]{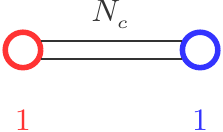}}\ \right) 
\label{eq:firstcase}  
\end{equation}

\medskip 

\medskip 

\subsection*{\texorpdfstring{$\bullet \ \ \text{5D} \ \  SU(2),\  N_{_f}=2$}{}}

\medskip 

\medskip 

The brane web is shown on the LHS of figure \ref{fig:22}. This time, there are two possible decompositions, \ref{fig:22}. Each one of them is mapped to a different magnetic quiver, \eqref{eq:neq}. The union of their CBs is equivalent to the HB of the original 5D quiver gauge theory we started from. The intersection of the two cones is given by a single symplectic leaf, as shown on the RHS of figure \ref{fig:22}.

\begin{figure}[ht!]    
\begin{center}  
\includegraphics[scale=0.6]{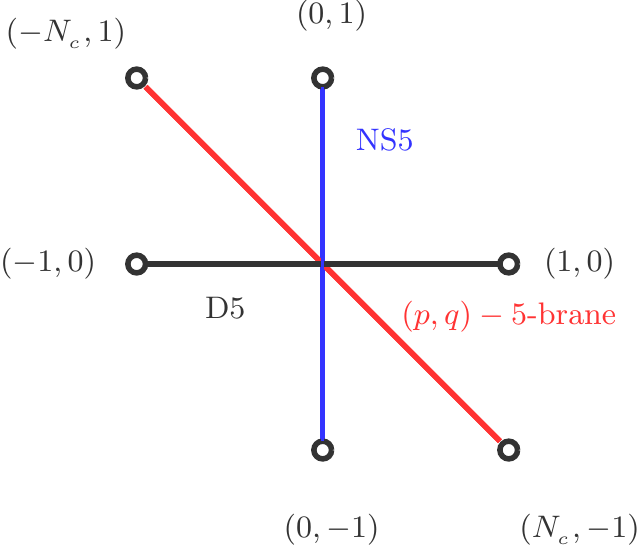} 
\qquad\qquad\qquad\qquad  
\includegraphics[scale=0.9]{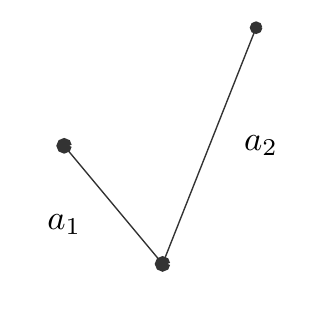} 
\end{center}
\end{figure}

\begin{figure}[ht!]    
\begin{minipage}[c]{1\textwidth}
\caption{\small 5-braneweb with added flavour (LHS) and HB Hasse diagram with two cones.}  
\label{fig:22}  
\end{minipage}   
\end{figure}

\begin{figure}[ht!]     
\begin{center}
\includegraphics[scale=0.6]{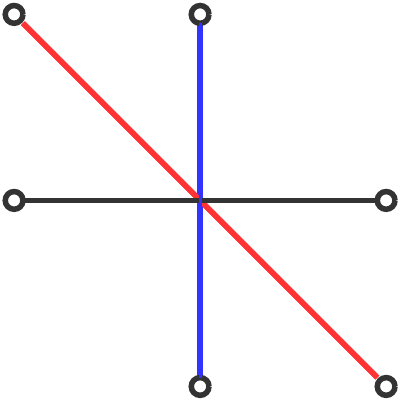}  
\ \ \ \ \ \ \ \ \ \ \ \ \ \ \ \ \ \ \ \ 
\includegraphics[scale=0.6]{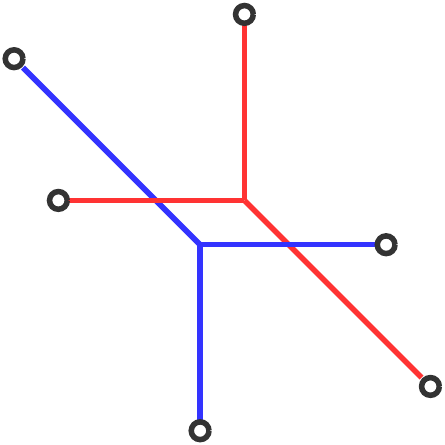}  
\caption{\small The two possible decompositions leading to the MQs featuring on the RHS of \eqref{eq:neq}.}  
\label{fig:2cones}  
\end{center} 
\end{figure}   

\begin{equation}  
\text{HB}_{_{\infty}}^ {\ \text{5D}}\ \left(\ \raisebox{-30pt}{\includegraphics[scale=0.75]{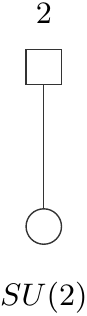}}\ \right) \ \equiv \ \text{CB}^{\ \text{3D}}\ \left(\ \raisebox{-23pt}{\includegraphics[scale=0.6]{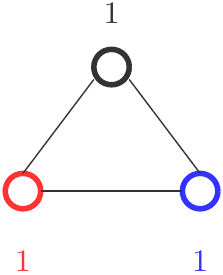}}\ \right) \ \cup \ \text{CB}^{\ \text{3D}}\ \left(\ \raisebox{-10pt}{\includegraphics[scale=0.6]{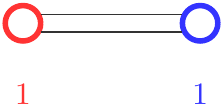}}\ \right) 
\label{eq:neq}
\end{equation}

\subsection*{\texorpdfstring{$\bullet \ \ 5D \ \  SU(2), \ N_{_f}=4$}{}}

\medskip 

\medskip 

\begin{figure}[ht!]
\begin{center} 
\includegraphics[scale=0.5]{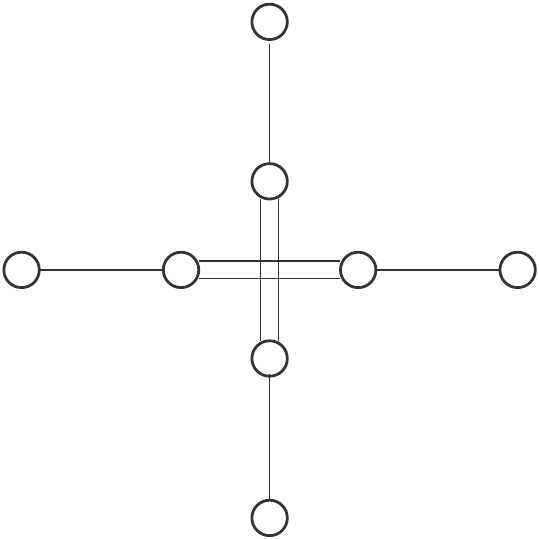} 
\ \ \ \ \ \ \ \ \ \ \ \ \ \ \ \ \ \ \ \ 
\includegraphics[scale=1]{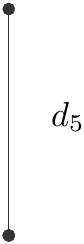} 
\end{center}
\end{figure}

\begin{figure}[ht!]
\begin{center} 
\caption{\small 5-braneweb and corresponding Hasse diagram for 5D $SU(2)$, $N_{_f}=4$.}  
\label{fig:11}  
\end{center} 
\end{figure}

\begin{equation}  
\text{HB}_{_{\infty}}^ {\ \text{5D}}\ \left(\ \raisebox{-30pt}{\includegraphics[scale=0.75]{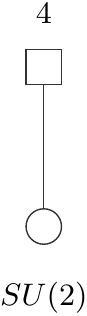}}\ \right) \ \equiv\ \text{CB}^{\ \text{3D}}\ \left(\ \raisebox{-25pt}{\includegraphics[scale=0.6]{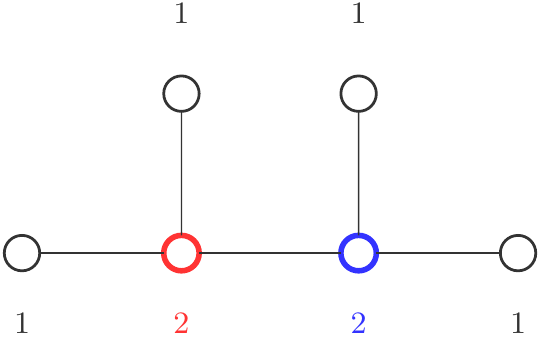}}\ \right) 
\end{equation}

\subsection*{\texorpdfstring{$\bullet \ \ 5D \ \ SU(3)_0, \ N_{_f}=6$}{}}

\begin{figure}[ht!]
\begin{center}
\includegraphics[scale=0.6]{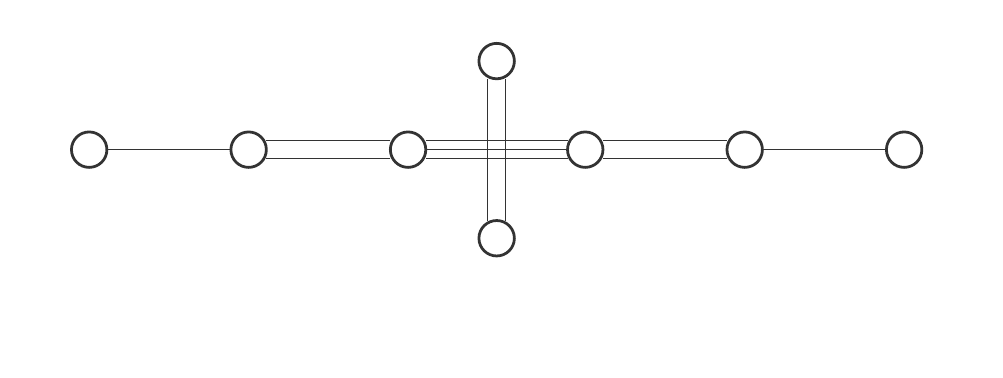}  
\qquad\qquad    
\includegraphics[scale=0.7]{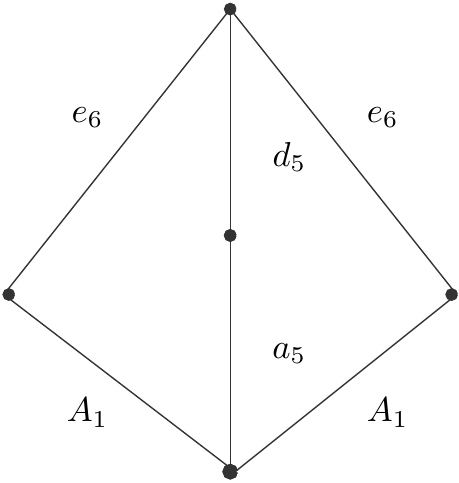}  
\end{center}
\end{figure}

\begin{figure}[ht!]    
\begin{minipage}[c]{1\textwidth}
\caption{\small 5-braneweb for 5D $SU(3)_0$, $N_{_f}=6$. The cone is unique since $k=0$.}  
\label{fig:44}  
\end{minipage} 
\end{figure}

\begin{equation}  
\text{HB}_{_{\infty}}^ {\ \text{5D}}\ \left(\ \raisebox{-30pt}{\includegraphics[scale=0.75]{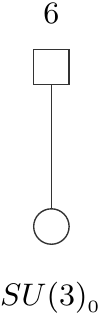}}\ \right) \ \equiv\ \text{CB}^{\ \text{3D}}\ \left(\ \raisebox{-25pt}{\includegraphics[scale=0.55]{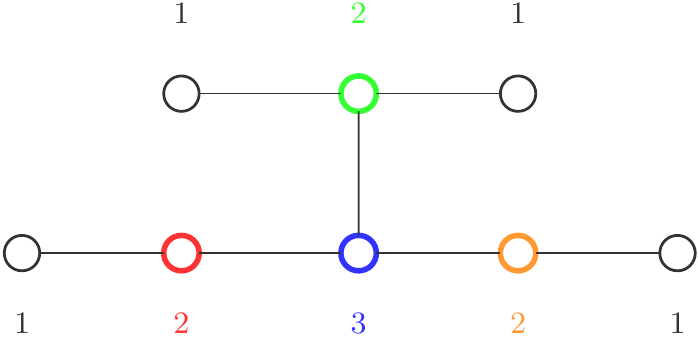}}\ \right) 
\end{equation}

\medskip 

\medskip 
\subsection*{\texorpdfstring{$\bullet\ \ 5D \ \ SU(6)_2, \ N_{_f}=8$}{}}  
\medskip 

\medskip 

\begin{figure}[ht!]     
\begin{center}
\includegraphics[scale=0.5]{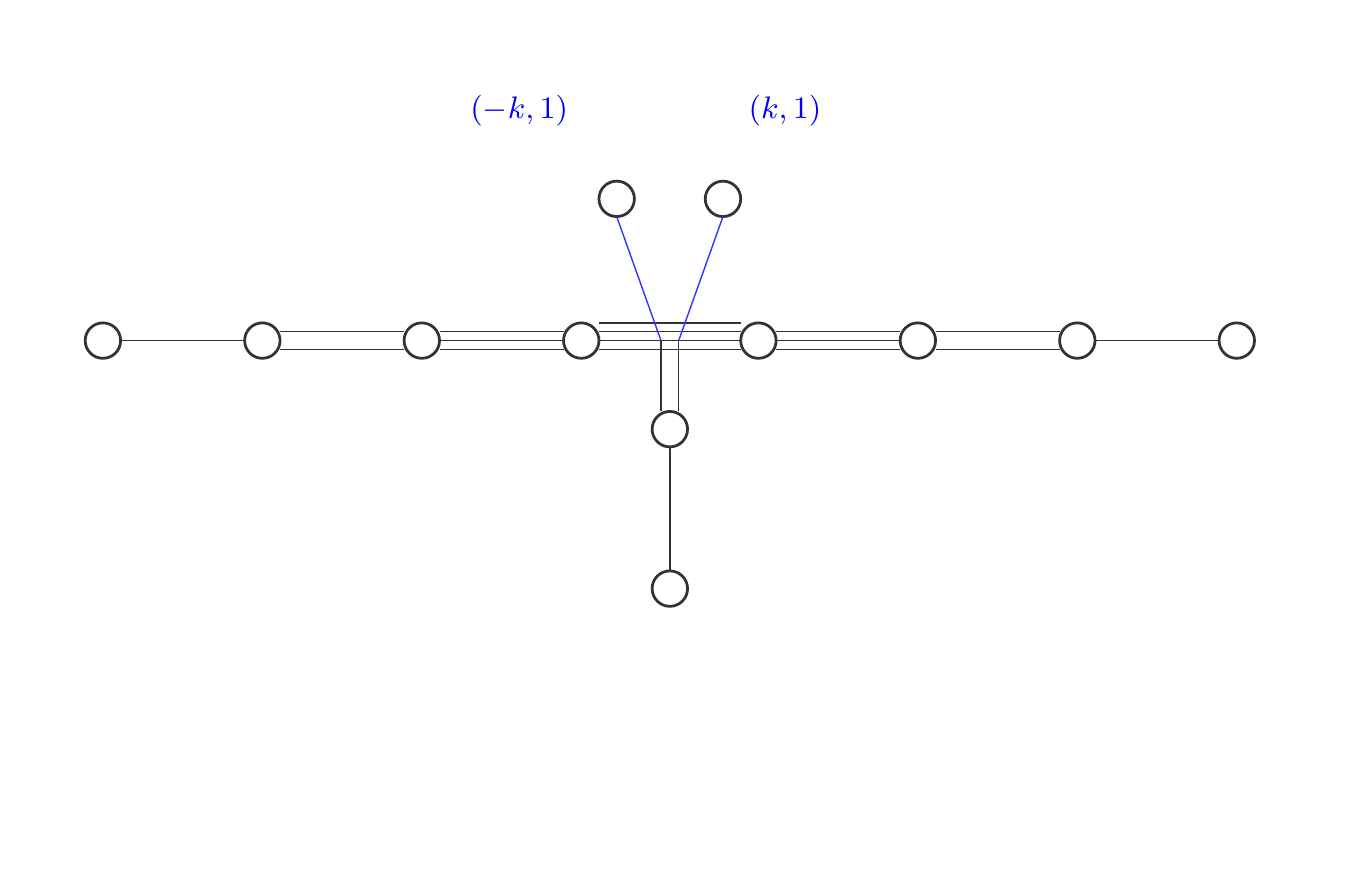}  \ \ \ \ \ \ \ \ \ \ \ \ 
\includegraphics[scale=0.6]{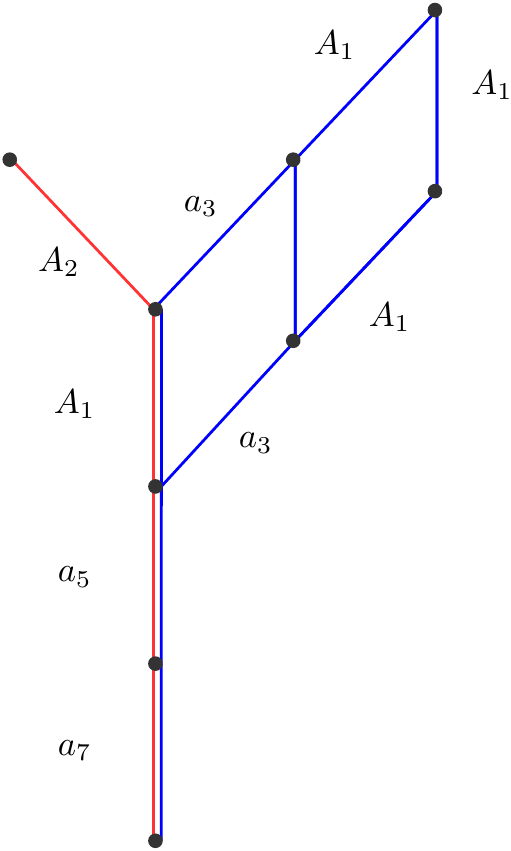}  
\caption{\small Adding a CS level $k\neq0$ leads the coexistence of two maximal decompositions of the 5-braneweb. The corresponding Hasse diagram therefore consists of 2 intersecting cones (denoted in red and blue on the RHS). These correspond to the mesonic and baryonic branches defining the overall HB. The intersection in this case goes along the vertical direction where the blue and red lines are parallel to each other.}  
\label{fig:33}  
\end{center} 
\end{figure}

\begin{equation}  
\text{HB}_{_{\infty}}^ {\ \text{5D}}\ \left(\ \raisebox{-30pt}{\includegraphics[scale=0.75]{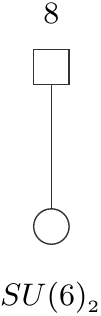}}\ \right) \ \equiv \ \text{CB}^{\ \text{3D}}\ \left(\ \raisebox{-25pt}{\includegraphics[scale=0.35]{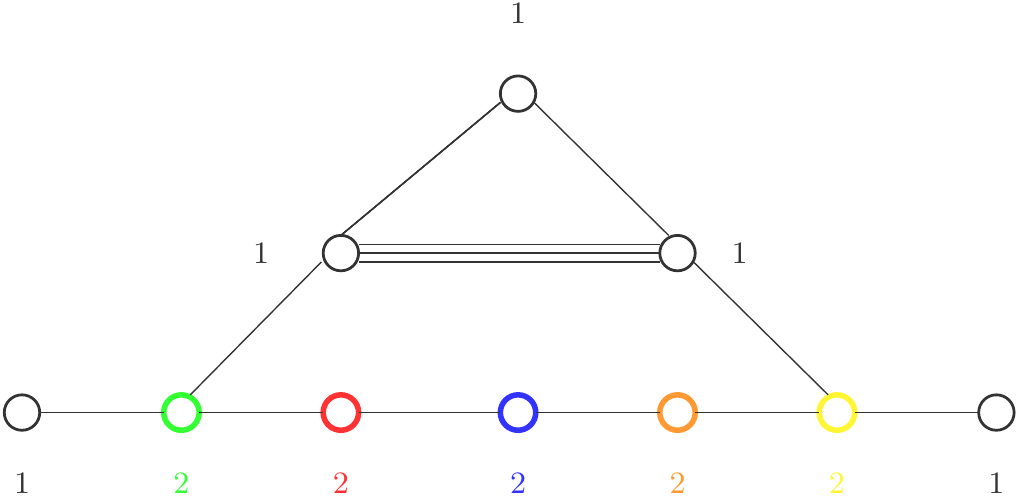}}\ \right) \ \cup \ 
\text{CB}^{\ \text{3D}}\ \left(\ \raisebox{-20pt}{\includegraphics[scale=0.35]{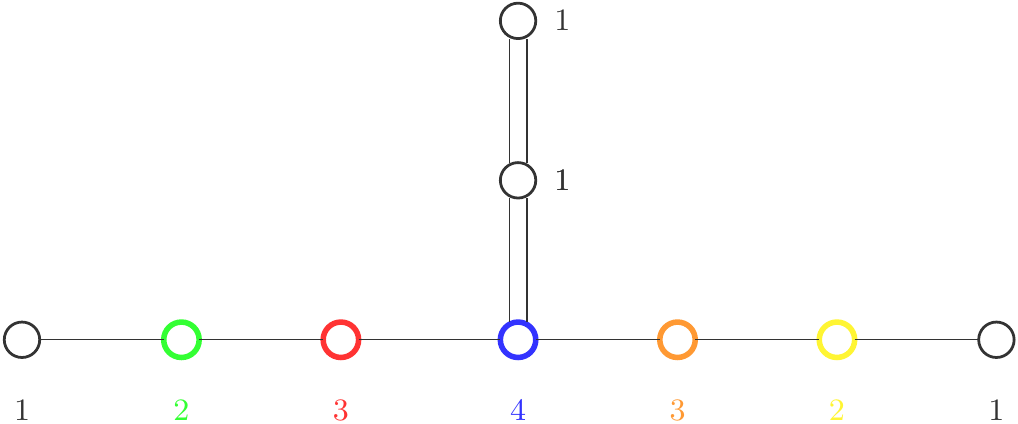}}\ \right) 
\end{equation}

From the point of view of the HB of quiver gauge theories with 8 supercharges in 5, 4, and 3 dimensions share a similar structure. In particular, the ramification of the HB due to the presence of multiple cones is preserved when considering the derived category of BPS objects in the lower-dimensional theory.

\subsection*{The 2-categorical structure of complete intersections}  

As previously mentioned, there are quiver gauge theories with 8 supercharges that do not admit a mirror symmetric 3D ${\cal N}=4$ theory description. The reason for this can be recast to the fact that its HB is not a single hyperk$\ddot{\text{a}}$ler cone, but, rather, the union of two, with nontrivial intersection, such as the cases shown in figure \ref{fig:22} and \ref{fig:33}. 

For classical gauge groups $G\equiv U(N), Sp(N)$ or $USp(2N), SO(2N)$, the moduli space is a complete intersection, meaning it is an algebraic variety whose dimension, $d$, is given by the following expression

\begin{equation}  
d\ \overset{def.}{=}\ g-r,     
\end{equation}  
where $g, r$ respectively denote the generators and relations between them. 

For the case in which a 3D ${\cal N}=4$ gauge theory, $T$, with given gauge group and suitable choice of matter content, admits a mirror dual theory, $T^{^{\text{V}}}$, the Hilbert series evaluated on the Coulomb branch (CB) of the former equals that of the Higgs branch (HB) of the latter,   

\begin{equation} 
\text{HS}\left[\text{CB}\  \left(T\right)\right]\ \equiv\ \text{HS}\left[\text{HB} \ \left(T^{^{\text{V}}}\right)\right],   
\label{eq:perfectmirror}   
\end{equation}     
hence   

\begin{equation}  
\text{HB}\left(\text{EQ}\right)\ \equiv\ \text{CB}\left(\text{MQ}\right). 
\label{eq:pm1}
\end{equation} 

Perfect realisation of ordinary mirror symmetry in \eqref{eq:perfectmirror} follows from the HB and CB of the dual theory both being hyperk$\ddot{\text{a}}$hler quotients. Recent developments in terms of magnetic quivers and Hasse diagrams highlighted the fact that setups where \eqref{eq:perfectmirror} effectively takes place involve Hasse diagrams featuring a single cone, such as in figure \ref{fig:11} and \ref{fig:44}. This feature naturally emerges from the HS calculation in the sense that \eqref{eq:HS1} would only account for the contribution from the generators of the unique cone involved in the construction of the diagram. On the other hand, for theories whose Hasse diagram is the union of two intersecting cones, \eqref{eq:HS} splits into terms associated to the generators of each individual cone, together with additional subtractive terms accounting for the nontrivial intersection in between them. 

Extended calculation of HS for arbitrary gauge theories with 8 supercharges has been the main focus of several recent works, mostly \cite{Bourget:2023cgs,Bourget:2019rtl,Ferlito:2016grh}. As shown in \cite{Bourget:2019rtl}, there are some cases in which one could trade a cone for an intersection, thereby indicating that the generators and the relations in between them encode the same information. This turns out to be of particular importance for our purposes, in particular in building the correspondence\footnote{A more detailed explanation of this will be provided in the next section, when comparing with \cite{Teleman:2014jaa}.} with the Drinfeld center\footnote{As stated in the introduction, the present work focuses on outlining the main proposal of connecting Drinfeld centers and magnetic quivers from a theoretical point of view, while reserving a more formal mathematical treatment of the same correspondence to a followup work by the same author, \cite{VP1}.}.   

Within the context of quiver gauge theories with 8 supercharges, the Hilbert series (HS) is a partition function counting chiral gauge-invariant operators, encoding the variety of vacua generated by such operators. The highest weight generating function (HWG), encodes the same information in a more succint way, that is more useful to be dealing with for cases involving higher rank, \cite{Ferlito:2016grh}.   

The general expression reads

\begin{equation}  
\begin{aligned}
\text{HS}_{_{N_{_{f}}}}(t;x_{_1},,...x_{_k})\ &\equiv\ \underset{j}{\sum}\ f_{_j}(x_{_1},,...x_{_k})\ t^{^j},
\end{aligned}   
\label{eq:HS1}   
\end{equation}
with $t$ denoting a fugacity for the highest weight of the $SU(2)_{_R}$ $R$-symmetry group providing the grading for the ring of functions, while $f_{_i}(x_{_1},,...x_{_k})$ are sums of characters for irreducible representations of the global symmetry group. From this follows the equivalence with \eqref{eq:hs} and \eqref{eq:hs1}. In the notation of \cite{Ferlito:2016grh}, this reads

\begin{equation}  
\begin{aligned}
\text{HS}_{_{N_{_{f}}}}(t;x_{_1},,...x_{_k})\ &\equiv\ \int d\mu_{_{G}}\ \text{HS}\left(\frac{\mathbb{C}[{\cal Q}, \tilde{\cal Q}]}{<\text{$F$-terms}>\ }\right),
\end{aligned}   
\label{eq:HS}   
\end{equation} 
with $M\ \overset{def.}{=}\ \tilde{\cal Q}{\cal Q}$ denoting the meson matrix.

For the case of 4D ${\cal N}=2$ SQCD with gauge group $G=SU(N_{_c})$ and $N_{_f}$ hypermultiplets in the fundamental representation of $SU(N_{_c})$, the calculation of \ref{eq:HS} requires: 

\begin{enumerate} 

\item solving the $F$-term equations, setting the derivatives of the superpotential to zero 

\item identifying the gauge-invariant operators. 

\end{enumerate}

For ${\cal N}=2$ SU(2) $N_{_f}=2$, the HS is a sum of rational functions of a quadruple of fugacities, $(t,z,x_{_1},x_{_2})$, \cite{Ferlito:2016grh}, where $t,z$ are associated to the $SU(2)_{_R}$ spin and $SU(2)$ gauge group, respectively, while $x_{_1},x_{_2}$ denote fugacities for the $SO(4)$ flavour symmetry,

\begin{equation}  
\begin{aligned}
\text{HS}(t;x_{_1},x_{_2})\ &\equiv\ \int d\mu_{_{SU(2)}}\ F(t,z,x_{_1},x_{_2})\\  
\ &\equiv\ \text{HS}(\mathbb{C}^{^2}/\mathbb{Z}_{_2};t,x_{_1})+\text{HS}(\mathbb{C}^{^2}/\mathbb{Z}_{_2};t,x_{_2})-1, 
\label{eq:hs3}
\end{aligned}
\end{equation}
with the first two terms corresponding to the two cones and the last term denoting the intersection in between them at the origin, as shown in figure \ref{fig:22}. When expanding it in terms of powers of $t$, we get the Plethystic logarithm (PL),

\begin{equation}  
\begin{aligned}
\text{PL}(t;x_{_1},x_{_2})\ &\equiv\ ([2;0]+[0;2])t^{^2}-([2;2]+2[0;0])t^{^4}+...,
\label{eq:hs4}
\end{aligned}
\end{equation}
with $[;]$ denoting the characters of the corresponding representations of $SO(4)$.
The character corresponding to a certain representation can be encoded in the corresponding Dynkin label. We can therefore choose a set of fugacities, $\mu_{_1},\mu_{_2}$, to keep track of them, enabling us to rewrite \eqref{eq:hs3} and \eqref{eq:hs4} as a highest weight generating function 

\begin{equation} 
\text{HWG}\left(t;\mu_{_1},\mu_{_2}\right)\ \equiv\ \text{PE}\left[\left(\mu_{_1}^{^2}+\mu_{_2}^{^2}\right)t^{^2}-\mu_{_1}^{^2}\mu_{_2}^{^2}t^{^4}\right],
\end{equation}
with the first two terms corresponding to the generators of the two cones, and the last (subtractive) term to their intersection in the Hasse diagram (cf. the RHS of figure \ref{fig:22}).

For the purpose of our analysis, the crucial step leading to the decomposition in cones, \eqref{eq:hs3}, is the application of the so-called primary decomposition of an algebraic variety and its associated ideal, \cite{Bourget:2019rtl}. Indeed, as explained in such reference, the different cones defining the HB arise upon disentangling the $F$-term equations, namely decomposing the ideal of the original algebraic variety into an intersection of primary ideals. For the case at hand, there are 2 intersecting ideals\footnote{Cases with multiple cones have recently been addressed in \cite{Bourget:2023cgs}.}.

\section*{Key points} 

The main message of this first section can be summarised as follows: 

\begin{itemize} 

\item Quiver gauge theories with 8 supercharges are characterised by a structure fully specified by knowing the generators and intersections of the cones in their HB Hasse diagram. The latter is obtained by implementing quiver subtraction on MQs associated to maximal decompositions of the 5-brane webs associated to the electric quiver. As such, it admits a 2-categorical structure description.

\item The moment map plays a key role in ensuring the identification of the underlying MQs. The flatness condition it is required to satisfy defines the HS, hence the HWG function, \cite{DAlesio:2021hlp}. Combining the works of \cite{Cabrera:2019izd,Bourget:2019aer,Bourget:2021siw,Bourget:2023cgs,Bourget:2019rtl,Ferlito:2016grh}, this constitutes a sample realisation of homological mirror symmetry. 

\item In presence of more than one cone in the Hasse diagram, the identification of the MQs whose CBs' union equals the HB of the original theory, is a generalised statement of homological mirror symmetry.

\end{itemize}  

We wish to emphasise that, while most of the above statements have been individually addressed in the literature, to the best of our knowledge, their combined application towards identifying the Drinfeld center in connection to CBs constitutes an original approach, to which the following section is devoted.

\section{Drinfeld Centers from Magnetic Quivers}  \label{sec:3}

We now turn to address an issue highlighted in a previous work by the same author, \cite{Pasquarella:2023deo}. Prior to explaining how this can be solved combining MQs and Drinfeld centers\footnote{We will explain the meaning of the latter in due course.}, for completeness, we will first review the features of the setup of \cite{Pasquarella:2023deo} that are mostly pertinent to the present treatment\footnote{We refer the reader to the first section of \cite{Pasquarella:2023deo} for the essential higher-categorical structure basics we will be using throughout the present treatment. Additional tools are explained in due course where needed.}, suitably incorporating elements of section \ref{sec:DCFMQS}. This section is thereby structured into three parts:   

\begin{itemize} 

\item Relying upon the description of the HB and CBs as algebraic varieties (cf. section \ref{sec:2.2}), we explain how gauging-by-condensation can be related to the poset ordering leading to the construction of Hasse diagrams, thanks to the unifying role of the moment map.

\item  We then turn to explaining how the identification of such moment map ensures the quiver gauge theory enjoys a generalisation of homological mirror symmetry, with the latter corresponding to the presence of a Drinfeld center, and an associated fiber functor for a certain 2-categorical structure, related to Rozansky-Witten theory, \cite{Rozansky:1996bq}.

\item We conclude opening a connection between the topics outlined in the present work and those of \cite{Teleman:2014jaa, CT}.

\end{itemize}  

The relation between posets and complex cohomology, already addressed in the literature, will be explored in greater mathematical detail in relation to identifying Drinfeld centers in a forthcoming work by the same author, \cite{VP1}. Such relation is intrinsic to the present treatment, thanks to the connections outlined in section \ref{sec:DCFMQS}; however, we are confident that a more mathematical treatment could also lead to further interesting insights that are worth exploring.

\subsection{Gauging-by-condensation and Hasse diagrams}  \label{sec:3.1}

We first start by recalling the configuration of interest in \cite{Pasquarella:2023deo}, motivating the importance of the Drinfeld center in geometric representation theory, \cite{CT}.

\subsection*{Relative and absolute QFTs}  

\begin{figure}[ht!]  
\begin{center}
\includegraphics[scale=0.75]{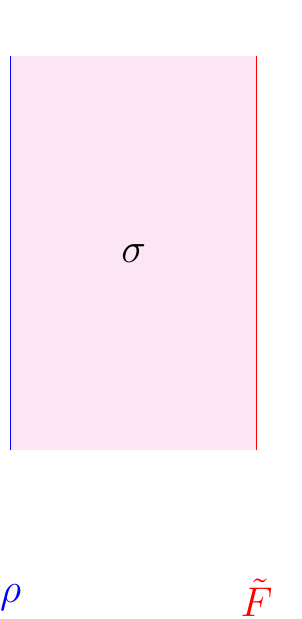} 
\ \ \ \ \ \ \ \ 
\includegraphics[scale=0.75]{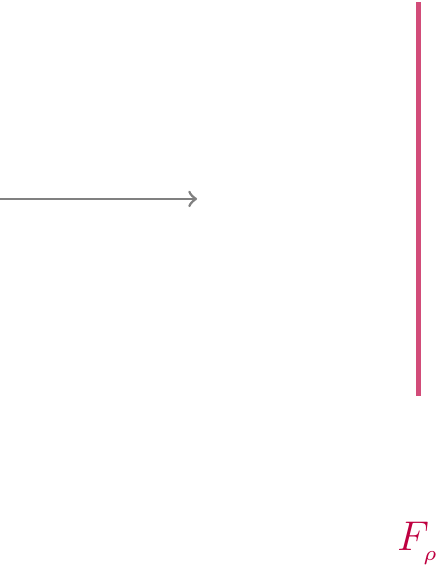}  
\caption{\small The Freed-Moore-Teleman setup, with $\tilde F$ denoting a relative QFT. Specifying the topological data $(\sigma,\rho)$, the resulting theory, $\tilde F_{_{\rho}}$ is absolute.}
\label{fig:FMT}    
\end{center}  
\end{figure}

In the formulation of \cite{Freed:2012bs}, a \emph{relative} field theory, $\tilde F$, requires additional topological data in order to be fully specified. Such data is encoded in a pair $(\sigma,\rho)$, referred to as \emph{quiche}. $\sigma$ is the \emph{symmetry topological field theory} (SymTFT), whereas $\rho$ geometrises the choice of boundary conditions for the fields defining the relative theory $\tilde F$. The overall system, depicted in figure \ref{fig:FMT}, gives rise to an \emph{absolute} QFT, $F_{_{\rho}}$.

Mathematically, the description outlined above can be formulated in terms of bordism in the following way. Fixing $N\in\mathbb{Z}^{^{\ge 0}}$, a quiche is a pair $(\sigma, \rho)$ in which $\sigma:\text{Bord}_{_{N+1}}(F)\rightarrow{\cal C}$ is an $N+1$-dimensional TFT, and $\rho$ is a right topological $\sigma$-module. The quiche is $N$-dimensional, hence it shares the same dimensionality as the theory on which it acts. Let $F$ be an $N$-dimensional field theory. A $(\sigma,\rho)$-\emph{module structure} on $F$ is a pair $(\tilde F, \theta)$, in which $\tilde F$ is a left $\sigma$-module and $\theta$ is an isomorphism 

\begin{equation}   
\theta\ : \rho\ \otimes_{_{\sigma}}\ \tilde F\xrightarrow{\ \ \simeq\ \ }\ F_{_{\rho}},  
\label{eq:LHS}  
\end{equation}   
of absolute $N$-dimensional theories, with \eqref{eq:LHS} defining the dimensional reduction leading to the absolute theory. $\sigma$ needs only be a \emph{once-categorified} $N$-dimensional theory, whereas $\rho$ and $\tilde F$ are relative theories.

\subsection*{Fiber functors}  

\begin{figure}[ht!]   
\begin{center}
\includegraphics[scale=0.6]{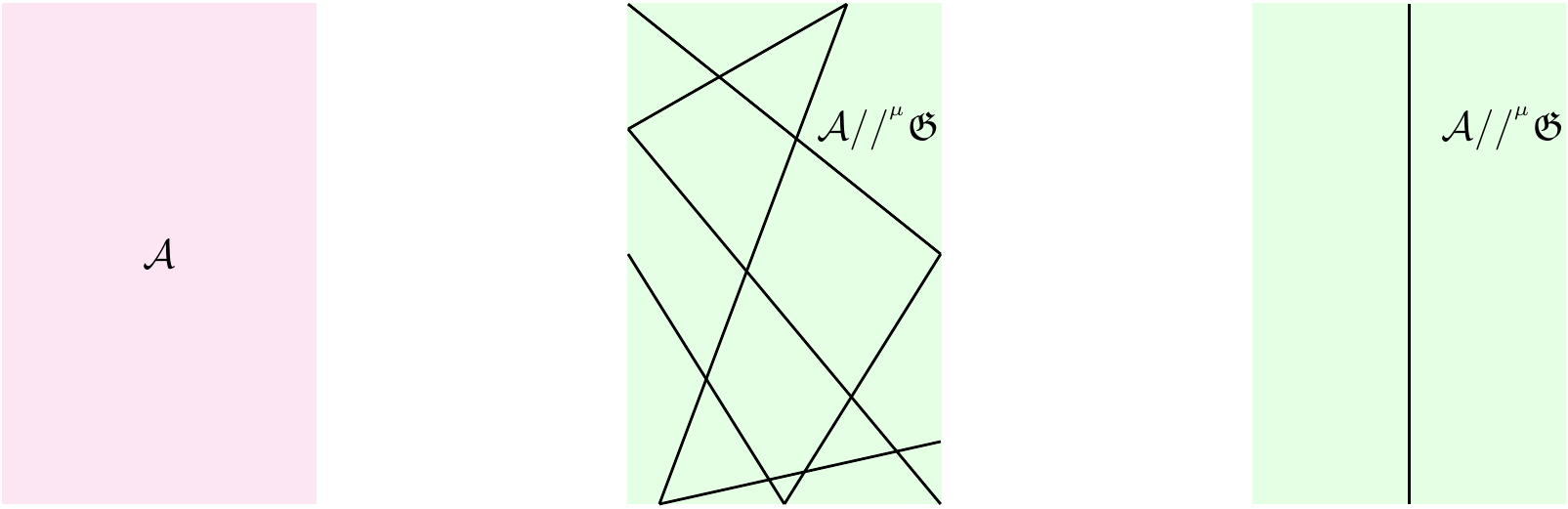}     
\caption{\small Gauging corresponds to condensing an algebra in a TFT. Idempotency ensures the resulting theory can be effectively thought of as featuring a unique defect, as shown on the RHS.}
\label{fig:codensationTJF}  
\end{center} 
\end{figure}

We now briefly overview what gauging a categorical structure actually means. In doing so, we refer to the work of many experts in the field, \cite{Gaiotto:2020iye,Kong:2020cie,Kong:2022cpy,Kong:2019byq,Kong:2019cuu,Kong:lastbutone,Gaiotto:2019xmp,Johnson-Freyd:2021tbq,Johnson-Freyd:2020usu,MYu,Freed:2012bs,Freed:2022qnc}, and, in particular \cite{TJF}. As explained in such reference, for any fusion n-category $\mathfrak{G}$, a fiber functor

\begin{equation} 
\boxed{\ \ \ \ {\cal F}:\ \mathfrak{G}\ \rightarrow\ \text{nVec}\color{white}\bigg]\ \ },  
\label{eq:functor}   
\end{equation}   
selects nVec as the image of a condensation algebra living in $\mathfrak{G}$, corresponding to a projection on the identity. The gauging process, can therefore be defined as a moment map\footnote{The choice of the same terminology as in section \ref{sec:2.1} is no coincidence, as will become clear in what follows.}

\begin{equation} 
\boxed{\ \ \ \ \mu:\ \mathfrak{G}\ \rightarrow \ {\cal A}\color{white}\bigg]\ \ }, 
\label{eq:mu}   
\end{equation}   
with ${\cal A}$ the algebra of invertible topological operators in $\mathfrak{G}$. Given \eqref{eq:mu}, the norm element

\begin{equation} 
N\overset{def}{=}\bigoplus_{g\in\mathfrak{G}}\ \mu(g)     
\label{eq:idemp}   
\end{equation}    
carries the structure of an n-categorical idempotent, also known as \emph{condensation algebra}, depicted in black in figure \ref{fig:codensationTJF}.

The requirement for \eqref{eq:idemp} to be a higher-idempotent is needed to ensure the flooding doesn't depend on the specific features of the network being adopted to perform the gauging. The algebra of topological operators that are left are denoted by ${\cal A}//^{^{\mu}}\mathfrak{G}$. The equivalence of the second and third picture from the left in figure \ref{fig:codensationTJF} follows from $N$ being a higher-condensation algebra. This pattern emerges when gauging the bulk SymTFT, where the objects of a certain higher-category, charged under a higher-form symmetry.

\begin{figure}[ht!]   
\begin{center}
\includegraphics[scale=0.6]{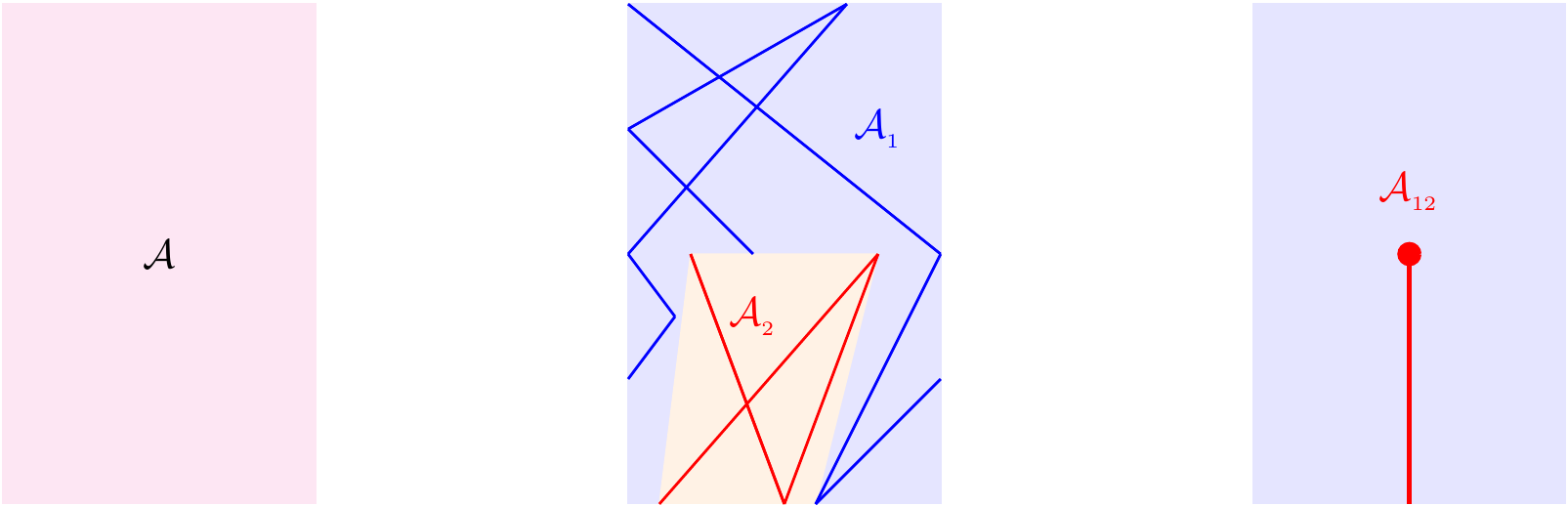}     
\caption{\small Condensing two different subalgebras, ${\cal A}_{_1}, {\cal A}_{_2}\ \subset\ {\cal A}$, the resulting theory corresponds to one with a changed phase with a condesating defect resulting from a relative condensable algebra, ${\cal A}_{_{12}}$ ending in the bulk. The defect at the endpoint is nontrivial, and can therefore be thought of as a Hom$(\mathbb{1}_{_{{\cal C}}},{\cal A}_{_{12}})$.}  
\label{fig:2codensationsTJF}  
\end{center} 
\end{figure} 

\begin{figure}[ht!]   
\begin{center}
\includegraphics[scale=0.7]{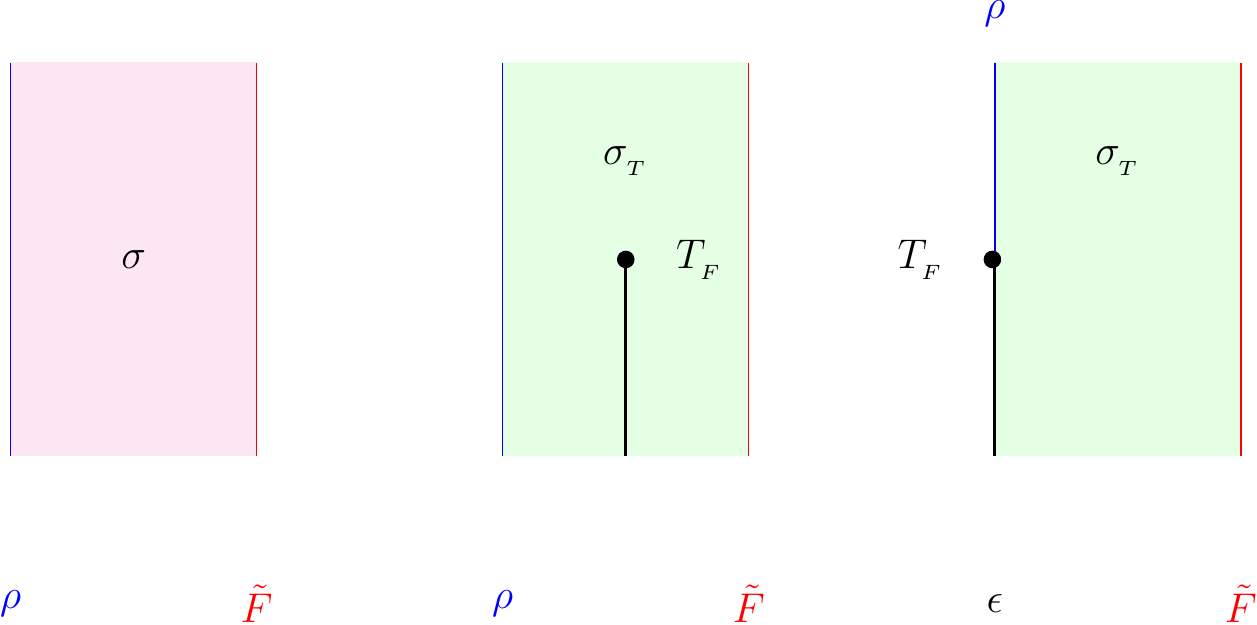}  \ \ \ \ \ \ \ \ \ \ \ \ \ \ \ \ \ \ 
\includegraphics[scale=0.8]{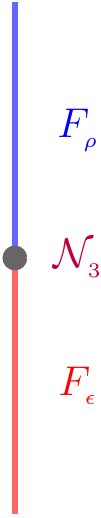} 
\caption{\small Adaptation of Freed-Moore-Teleman to the case involving twisted condensation defects. The figure on the far right corresponds to the case of interest to us, namely a configuration involving two different absolute 4D gauge theories separated by a defect. As we shall see, such defect is intrinsically non-invertible, corresponding to the presence of a relative uncondensed subalgebra, dressing ${\cal N}_{_3}$ in a nontrivial way. In higher-categorical terms, it corresponds to a fusion tensor category implementing the morphisms between the operators charged under the gauged symmetry.}
\label{fig:FMTE}  
\end{center} 
\end{figure}

\begin{figure}[ht!]     
\begin{center}
\includegraphics[scale=1]{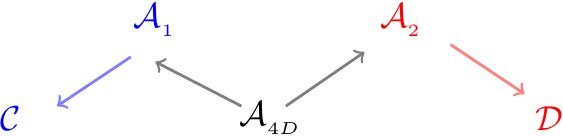}  
\caption{\small The composition of functors leading to the relative condensable algebra \eqref{eq:compfunct2}. Its derivation is explained in \cite{Pasquarella:2023deo}. Here ${\cal C},{\cal D}$ are the 2-categories characterising two different absolute theories separated by an intrinsically non-invertible defect dressed by the relative condensable algebra $_{_{{\cal A}_{_1}}}{\cal A}_{_{{\cal A}_{_2}}}$.}  
\label{fig:a1aa2}  
\end{center} 
\end{figure}

In our previous work, \cite{Pasquarella:2023deo}, we already argued that, unlike figure \ref{fig:codensationTJF}, the third picture from the left in figure \ref{fig:FMTE} does not admit a straightforward expression for the fiber functor as \eqref{eq:functor}. Nevertheless, we claimed that a composite fiber functor could still be assigned thanks to the identification of the underlying algebraic structure of the composite theories

\begin{equation}
\boxed{ \color{white}{bla} \color{black} _{_{{\cal A}_{_1}}}{\cal A}_{_{{\cal A}_{_2}}}\ \overset{def.}{\equiv}\ {\cal A}_{_1}\ \otimes_{_{\cal A}}\ {\cal A}_{_2}\ \ \color{white}\bigg]\  },  
\label{eq:compfunct2}   
\end{equation}   
resulting from the composition drawn in figure \ref  {fig:a1aa2}.

Our proposal in \cite{Pasquarella:2023deo} was that the corresponding fiber functor should really be defining the partition function of a 3D theory, whose identification we anticipated being the core topic of the present work. The remainder of this section is devoted to prove that the resulting 3D theory is a 3D ${\cal N}=4$ quiver gauge theory whose HB, when viewed as a complete intersection algebraic variety, enables to define the Drinfeld center of its underlying 2-categorical structure. In doing so, we need to completely specify its 1- and 2-morphisms, namely the generators and the relations in between them, both needed in the evaluation of the HB Hilbert series. As explained in section \ref{sec:DCFMQS}, the HS is nothing but the sum of gauge-invariant algebraic operators living in the chiral ring of the theory.

The Drinfeld center is defined as the 0$^{th}$-Hochschild cohomology, namely the endomorphisms of the identity element of a given category, ${\cal C}$:

\begin{equation}   
\boxed{\color{white}{blan} \color{black}\mathfrak{Z}({\cal C})\ \color{black}\overset{def.}{=}\ \text{End}(\mathbb{1}_{_{{\cal C}}})\ \equiv\ \text{Hom}(\mathbb{1}_{_{{\cal C}}},\mathbb{1}_{_{{\cal C}}})\ \equiv\ HH^{^{{\bullet}\  0}}({\cal C})\ \  \color{white}\bigg]\  }.
\end{equation}

The existence of the Drinfeld center and that of the fiber functor are mutually guaranteed, since, by definition, 

\begin{equation}   
\boxed{\color{white}{blan}  \color{black}{\cal F}:\ \mathfrak{Z}({\cal C})\ \longrightarrow\ {\cal C}\ \ \color{white}\bigg]\  }. 
\end{equation}

From what we have just outlined, it clearly follows that, the issue of identifying the Drinfeld center is equivalent to that of assigning a fiber functor projecting to the category ${\cal C}$. For the purpose of our treatment, namely building connection with geometric representation theory, ${\cal C}$ is really meant to be the category of representations, with the latter being described in terms of 2D TFTs, namely Lagrangian submanifolds living in a symplectic manifold $X$, playing the role of boundary conditions for 3D Rozansky-Witten theory, $RW_{_X}$, \cite{Rozansky:1996bq}. The meaning of this will be explained in section \ref{sec:homms} and \ref{sec:3.3}.

\subsection*{Relation to Hasse diagrams} 

The relation in between gauging-by-condensation and the construction of Hasse diagrams basically follows from the underlying role played by the moment map. In the former, it comes hand-in-hand with the definition of the fiber functor, meaning a partition function can be assigned to the resulting absolute theory, whereas, in the context of quiver gauge theories, flatness of the moment map ensures a Hilbert series can be fully specified by the primitive ideals and their mutual intersections. Importantly, the poset structure of the Hasse diagram admits a complex cochain description, thereby opening the possibility of studying this from a more mathematically rigorous point of view. We will be reporting on this in an upcoming work, \cite{VP1}, showing connection with \cite{Teleman:2014jaa,Gonzalez:2023jur} in greater detail. 

As explicitly argued in the previous section, the crucial assumption enabling us to relate magnetic quivers with the realisation of homological mirror symmetry is the assumption that the moduli space of the theory being dealt with is a complete intersection. Precisely thanks to this we could describe the emergence of a 2-categorical structure, whose 1- and 2-morphisms are the generators of the cones featuring in the corresponding Hasse diagram, and their intersection, respectively. Crucially, agreement in between the electric and magnetic calculation of the Hilbert series requires specifying the primitive ideals and their intersections. As explained in section \ref{sec:2.2}, to each primitive ideal corresponds a different cone within the HB of the electric theory. Importantly, the full HB needs to account for the mutual intersection of the cones involved in the construction of the Hasse diagram, hence their intersections can in turn be related to an algebraic variety with an associated ideal resulting from the intersection of ideals defining different cones.

Having said this, we therefore conclude that figure \ref{fig:a1aa2} encodes the same information as a Hasse diagram with two intersecting cones, such as on the one on the RHS of figure \ref{fig:22}.

\subsection{Drinfeld centers and mirror symmetry}  \label{sec:homms}

As a first key result of our work, we therefore wish to highlight the following statement as being applicable to the theories specified in section \ref{sec:DCFMQS}, namely quiver gauge theories with 8 supercharges, 
   \medskip    
   \medskip
\color{blue}

\noindent\fbox{%
    \parbox{\textwidth}{% 
   \medskip    
   \medskip
   \begin{minipage}{20pt}
        \ \ \ \ 
        \end{minipage}
        \begin{minipage}{380pt}
      \color{black}  \underline{Key point:} identifying the Drinfeld center is equivalent to assigning a Hilbert series with irreducible representations, i.e. specifying the cones in the Hasse diagram, and their intersection.
        \end{minipage}   
         \medskip    
   \medskip
        \\
    }%
}
 \medskip    
   \medskip     \color{black}

The remainder of the present section is devoted to proving such assertion. In doing so, we mostly refer to \cite{Teleman:2014jaa, CT}, which was one of the first motivations of this work. A more detailed mathematical description of the correspondence with such references, as well as with \cite{Gonzalez:2023jur}, is the core topic of an upcoming work by the same author, \cite{VP1}. For the moment, we wish to highlight that such a correspondence exists, and can be understood from a mathematical physicist's perspective. As a bi-product, we show how magnetic quivers can be combined with higher categories to extend the setup of  \cite{Teleman:2014jaa, CT} for describing higher dimensional QFTs admitting a quiver gauge theory description.   

Importantly, we emphasised at different points of this treatment the crucial assumption of the moduli space being a complete intersection. Indeed, the case of non-complete intersections requires going beyond the Drinfeld center, and turning to higher Hochschild cohomologies, \cite{CT}. In \cite{VP1} we aim to address the adaptation of the present treatment to such cases by developing a unifying mathematical language bridging the two procedures. In the interest of this, in section \ref{sec:3.3} we will set the stage for the more detailed analysis of \cite{VP1}.

Prior to doing so, we will: 

\begin{enumerate} 
\item First explain  the notionof 2-fiber products, needed for the proof of our main statement. 

\item Then, we will briefly recapitulate the notion of homological mirror symmetry, introduced in section \ref{sec:2.1} as a correspondence in between categories of representations. 

\end{enumerate}

As we shall see, they are both play a crucial role within the context of geometric representation theory, with 2D TQFTs generalising the notion of cohomology, \cite{Teleman:2014jaa}.

\subsection*{2-fiber products} 

This brief digression is meant to provide some preliminary mathematical tools needed for proving our main result. If $\cal C$ is a 2-category of categories, then there is a notion of a 2-fiber product, $\cal C$$_{_1}\ \times_{_{\cal D}}\ \cal C$$_{_2}$, also denoted by a diagram 

\begin{equation}  
{\cal C}_{_{1}}\ \xrightarrow{f}\ {\cal D}\ \xleftarrow{g}\ {\cal C}_{_2}  
\end{equation}   
is the category of triples $(c_{_1}, c_{_2},\phi)$, where $c_{_1}\in{\cal C}_{_1}$, $c_{_2}\in{\cal C}_{_2}$, and $\phi:f(c_{_1})\ \xrightarrow{\sim}\ g(c_{_2})$ is an isomorphism of their images $\cal D$. For an object $P\in\cal C$, a 2-fiber product $X$$\ \times_{_S}\ $$Y$ associated to the diagram 

\begin{equation}  
X\ \xrightarrow{f}\ S\ \xleftarrow{g}\ Y  
\end{equation} 
is a quadruple $(P,p,q,\phi)$, with 1-morphisms $p:P\rightarrow X, q:P\rightarrow Y$, and a 2-isomorphism

\begin{equation}  
\phi: f\ \circ\ p\ \simeq\ g\ \circ\ q  
\end{equation}  
such that $\forall Z\in\cal C$, the natural functor     

\begin{equation} 
\boxed{\ \ \ \text{Hom}_{_{\cal C}}\left(Z,P\right)\ \rightarrow\ \text{Hom}_{_{\cal C}}\left(Z,X\right)\times_{_{\text{Hom}_{_{\cal C}}(Z,S)}}\ \text{Hom}_{_{\cal C}}(Z,Y) \color{white}\bigg] \ \ }    
\label{eq:fibermirror}   
\end{equation}    
is an equivalence of categories. As we shall see, this is basically equivalent to a statement of homological mirror symmetry adapted to the configuration depicted on the second picture from the right in figure \ref{fig:FMTE}.

\subsection*{Homological mirror symmetry}

 As previously explained in \cite{CT1}, 3D mirror symmetry can be rephrased in terms of topological representation theory thanks to the Drinfeld center, \cite{CT}. We now briefly overview such argument, in order to make contact with the setting we are dealing with.
 
\begin{figure}[ht!]  
\begin{center} 
\includegraphics[scale=0.5]{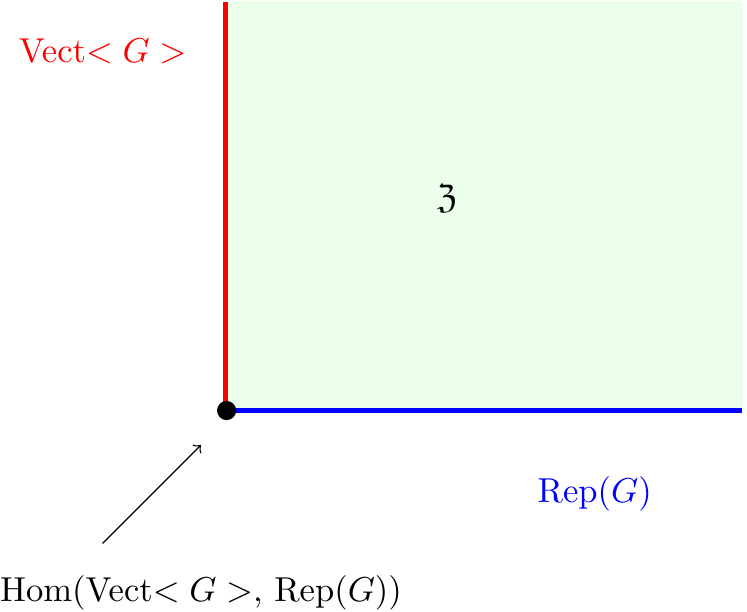} 
\ \ \ \ \ \ \ \ \ \ \ \ \ \ \ \ 
\includegraphics[scale=0.7]{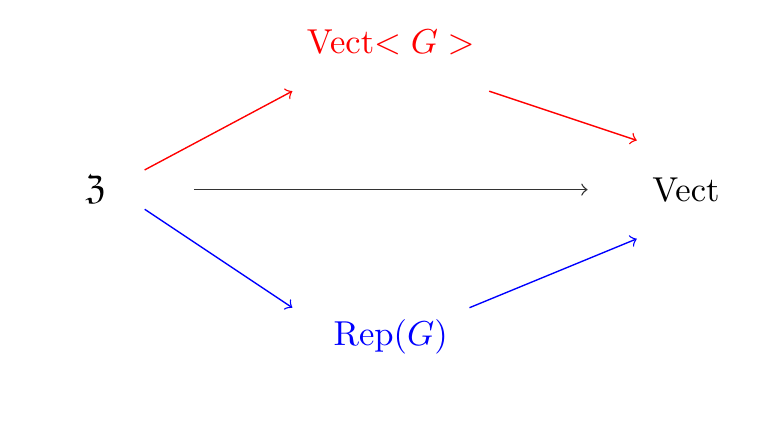} 
\caption{\small The black arrow in the diagram on the RHS denotes the fiber functor mapping from the Drinfeld center to the category of representations. Correspondingly, on the LHS, the Drinfeld center is the green shaded region whose boundaries are the regular and trivial representations, denoted in blue and red, respectively. For the case in which the two are related by ordinary mirror symmetry, their intersection contains the same information as $\mathfrak Z$, and therefore a fiber functor which is fully forgetful can be assigned.}
\label{fig:D}
\end{center}    
\end{figure}

In section \ref{sec:2.1} we explained that homological mirror symmetry consists in the proposed agreement between two categories, namely:

\begin{itemize}  

\item  Fukaya's $A_{_{\infty}}$-category, ${\cal F}(X)$, on the symplectic side,

\item and the derived category with Yoneda structure, $D^{^b}\mathfrak{Coh}(X^{^{\text{V}}})$, on the complex side, 

\end{itemize} 
with $X$ a symplectic manifold. A Fukaya category of a symplectic manifold, $X$, is an $A_{_{\infty}}$-category with Lagrangian submanifolds $L\ \subset\ X$ as objects, whose intersections define the Hom-space. For the case in which the two Lagrangians are related by mirror symmetry, such as is the case between Neumann and Dirichlet boundary conditions, the LHS of figure \ref{fig:D} is such that the intersection is able to account for the Drinfeld in a straightforward manner. This is what characterises a fully-extendable TQFT. 
The need for the Hom-space to be specified in order to fully determine the Drinfeld center at arbitrary distance w.r.t. the boundary conditions (i.e. the Lagrangian submanifolds), signals the presence of a non-invertible defect separating the resulting absolute theories, and therefore connects with the setup of figure \ref{fig:FMTE} involving double algebraic condensation. A generalisation of this will be outlined in section \ref{sec:3.3}, as shown in \cite{Teleman:2014jaa}. For the moment we outline an example proposed in \cite{CT1} to describe the realisation of 3D mirror symmetry in terms of category of representations. 

Take a finite group $G$ and two tensor categories Vect$<G>$ and Rep ($G$), with associated tensor products $*$ and $\otimes$. If the two tensor categories are Morita equivalent, namely,

\begin{equation} 
\left(\text{Vect}<G>,*\right)\ \equiv\ \left(\text{Rep}(G),\otimes\right)  
\end{equation} 
they share the same Drinfeld center, as shown on the LHS of figure \ref{fig:D}. What we have just said can be succintly rephrased as follows

\begin{equation} 
\boxed{ \color{white}{blac} \color{black} \text{Vect}<G>\ \otimes_{_{\mathfrak{Z}}}\ \text{Rep}(G)\ \equiv\ \text{Vect}\ \ \color{white}\bigg]\  },        
\label{eq:crucial}   
\end{equation}    
which is a particular realisation of \eqref{eq:fibermirror} for the case of a 2-category whose objects live in the Drinfeld center, whereas the 1-morphisms are the Lagrangian submanifolds living in the symplectic manifold associated to the theory of interest. We will come back to this more explicitly in section \ref{sec:3.3} when building connection with \cite{Teleman:2014jaa} in addressing Rozansky-Witten theory.

As shown on the RHS of figure \ref{fig:D}, \eqref{eq:crucial} corresponds to the definition of the category of representation associated to a fully forgetful and fully faithful fiber functor 

\begin{equation}   
\boxed{\color{white}{blan}  \color{black}{\cal F}:\ \mathfrak{Z}\ \longrightarrow\ \text{Vect}\ \ \color{white}\bigg]\  }. 
\label{eq:newf}  
\end{equation}  

For the case in which the two representations are associated to the choice of Dirichlet and Neumann boundary conditions, we know from their Symmetry TFT (SymTFT) realisation, that they can be placed on the topological boundary of the quiche, with a non-invertible defect separating them. Indeed, this configuration is precisely of the kind featuring on the far right of figure \ref{fig:FMTE}, for the case where the theory is coupled to a bulk SymTFT with a certain gauge group, and another with the fully-condensed condensable algebra\footnote{These statements refer to the condensable algebra specified by the moment map \eqref{eq:mu}.}.  

Hence, combined together, \eqref{eq:crucial} and \eqref{eq:newf} agree with the fact that \eqref{eq:compfunct2} can also be assigned a 2-fiber functor, built from the composition specified in the diagram \ref{fig:a1aa2}.

This completes the proof of our main claim in \cite{Pasquarella:2023deo}. Further supportive evidence is provided in the following subsection, where we will see the adaptation of the tools outlined in section \ref{sec:3.1} and \ref{sec:homms} to a 3D theory which is closely related to the supersymmetric theories addressed in section \ref{sec:2.2}.

\subsection{Drinfeld centers from magnetic quivers of 3D \texorpdfstring{${\cal N}=4$}{} gauge theories} \label{sec:3.3}   

In this concluding subsection, we combine the tools encountered throughout our treatment, drawing additional major conclusions, as well as setting the stage for further investigations, \cite{VP1}, where these topics will be addressed with a more mathematically-oriented approach, while always keeping track of the underlying physics such tools are meant to probe.

\subsection*{Rozansky-Witten Theory and the Drinfeld center}   

One of the examples encountered in section \ref{sec:2.2}, namely 5D ${\cal N}=1$ SQCD with $SU(2)$ with $N_{_f}=2$, is the simplest case involving two cones with nontrivial intersection in its Hasse diagram. As such, it is an example where ordinary mirror symmetry in the sense of \eqref{eq:perfectmirror} does not apply for the 3D ${\cal N}=4$ theory arising under dimensional reduction. Nevertheless, a suitable mirror dual can still be assigned once the generators of the two cones featuring in the Hasse diagram, together with their intersection, are correctly accounted for in the calculation of the Hilbert series (HS), i.e. the partition function counting the gauge-invariant operators living in the chiral ring.  
In the previous section we explained how the magnetic quiver prescription enables to identify the HS correctly for cases where ordinary mirror symmetry does not apply.

To make contact with the calculation of Drinfeld centers described in section \ref{sec:3.1} and \ref{sec:homms}, it is instructive to consider the case\footnote{Notice that this precisely falls within the case of quiver gauge theories with 8 supercharges to which the magnetic quiver and Hasse diagram prescription of section \ref{sec:2.2} applies.} of 4D $SU(2)$ ${\cal N}=2$, first studied by Seiberg and Witten within the context of its 3D reduction, \cite{Seiberg:1996nz}. Its low-energy limit is described as a sigma-model in the space of vacua, and was then identified with the Rozansky-Witten theory (RW), \cite{Rozansky:1996bq}, of the hyper-kah$\ddot{\text{a}}$ler Atiyah-Hitchin manifold. Later developments of this by Kapustin, Rozansky, \cite{Kapustin:2009uw}, and later with Saulina, \cite{Kapustin:2008sc}, addressed the 2-category of branes of such 3D theories, with the latter containing smooth holomorphic Lagrangians, $L$, whose definition was provided in section \ref{sec:2.1} when introducing Fukaya categories associated to symplectic manifolds, and later recalled in section \ref{sec:homms}. For our purposes, it is particularly useful to describe this setup as a 2-categorical structure of the kind we have been emphasising throughout the entire treatment\footnote{In the remainder of the present work we will denote the objects of such 2-category, namely the branes living in the Drinfeld center, simply by KRS, i.e. the initials of the authors of \cite{Kapustin:2008sc}. These are the objects whose description one needs to ensure agreement of under mirror symmetry.}. In particular, Lagrangian submanifolds will be 1-morphisms specifying the boundary conditions of objects living in the 3D RW$_{_{X}}$ theory associated to a given symplectic manifold $X$.

Concretely, in the RW model with a given gauge group $G$, one must geometrically describe two functors from the 2-category of linear categories with $G$-action to linear categories: 

\begin{enumerate}  

\item the forgetful functor, keeping track of the underlying category (describing the pre-gauged TQFT), associated with the regular representation of $G$,

\item the functor mapping to the invariant category, generating the gauged TQFT, associated with the trivial representation of $G$. 

\end{enumerate}   

As explained in section \ref{sec:homms}, this already encodes the idea of mirror symmetry, whose generalisation is realised thanks to the composite fiber functor described in \cite{Teleman:2014jaa} (and overviewed in the concluding part of this section), sharing the same2-fiber product structure of equations \eqref{eq:compfunct2}, \eqref{eq:fibermirror} and \eqref{eq:crucial}. This further supports the statement that the realisation of homological mirror symmetry is mapped to an agreement of the calculation of invariants and Hilbert series as sums over characters. 

\begin{figure}[ht!]     
\begin{center}
\includegraphics[scale=0.75]{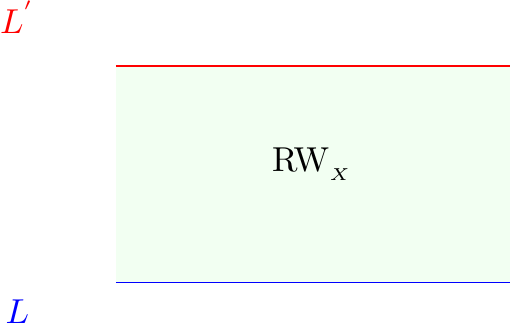}  
\ \ \ \ \ \ \ \    \ \ \ \ \ \ \ \ \ 
\includegraphics[scale=0.75]{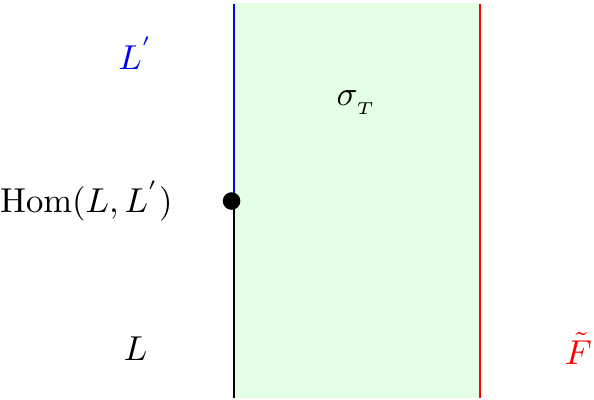}  
\caption{\small Adaptation of the RW$_{_X}$ theory (on the LHS) to the quiche of a relative field theory (RHS).}  
\label{fig:RW}  
\end{center} 
\end{figure}

Geometrically, such agreement is expected to show in the description of objects living in the bulk of RW theory. The objects in the 2-category of KRS are branes in the BFM space, \cite{Teleman:2014jaa,BFM}, whose boundaries are Lagrangians $L,L^{^{\prime}}\ \subset\ X$, $X$ a symplectic manifold. As shown on the LHS of figure \ref{fig:RW}, $L,L^{^{\prime}}$ are 1-morphisms generating the theory upon acting on the trivial theory   

\begin{equation}  
\boxed{\ \ \  L,L^{^\prime}:\ \text{Id}\ \longrightarrow\ \text{RW}_{_X}\color{white}\bigg]\ \ }.    
\end{equation}   
whereas Hom$(L,L^{^{\prime}})$ denotes the 2-morphisms between them.

For the case in which $L, L^{^{\prime}}$ are related by ordinary mirror symmetry, the 2-morphisms encode all the information, and the 2-category can be completely reconstructed from them. This case corresponds to the so-called fully-extended TQFTs. However, for cases involving a generalised notion of mirror symmetry, decategorification might lose track of important information in the original theory one is attempting to describe. Practically, this means we need to specify, both, the 1-morphisms and the 2-morphisms to reconstruct the bulk of the BFM space, and completely describe the KRS branes living in them.

In connection with section \ref{sec:homms} and our previous work, \cite{Pasquarella:2023deo}, it is instructive to visualise what we have just said as shown in figure \ref{fig:RW}. Fully specifying the topological boundary conditions leads to an absolute theory. In case Hom$(L,L^{^\prime})$ leads to a non-invertible defect, the fiber functor, hence the Drinfeld center associated to figure \ref{fig:RW}, needs a 2-fiber product structure to be fully specified, as also explained in \cite{Pasquarella:2023deo}. In particular, this highlights the importance of specifying the entire underlying 2-categorical structure, namely the Lagrangian submanisfolds as well as the homomorphisms between them, ultimately ensuring the objects in the 2-category of KRS, can be fully specified. From the arguments outlined in the preliminary part of this work, we are therefore led to conclude that the configuration on the RHS of figure \ref{fig:RW} is dual to another theory with the same Drinfeld center and only one Lagrangian submanifold on the topological boundary condition. From section \ref{sec:homms}, this provides a sample realisation of generalised homological mirror symmetry.

%The regular representations are categories of coherent sheaves ...  

\subsection*{Symplectically induced representations}

In conclusion, we outline more explicitly the connection with the work of \cite{Teleman:2014jaa}, thereby setting the stage for a more detailed mathematical treatment, which is the main focus of an upcoming work by the same author, \cite{VP1}. Importantly, \cite{Teleman:2014jaa} addresses the question from the point of view of derived representations. In particular, in such reference, it is shown that, in some cases, for homological mirror symmetry to be realised, one needs to be able to account for higher Hochschild cohomologies. For the examples we addressed in the present work, we were able to determine the Drinfeld center as fully specifying the invariants of the underlying 2-categorical structure. However, we believe that under the more mathematical description of \cite{VP1} we will be able to address the issue of going to higher Hochschild cohomologies, and seeing what this maps to from the point of view of supersymmetric quiver gauge theories. 

\begin{figure}[ht!]     
\begin{center}
\includegraphics[scale=0.75]{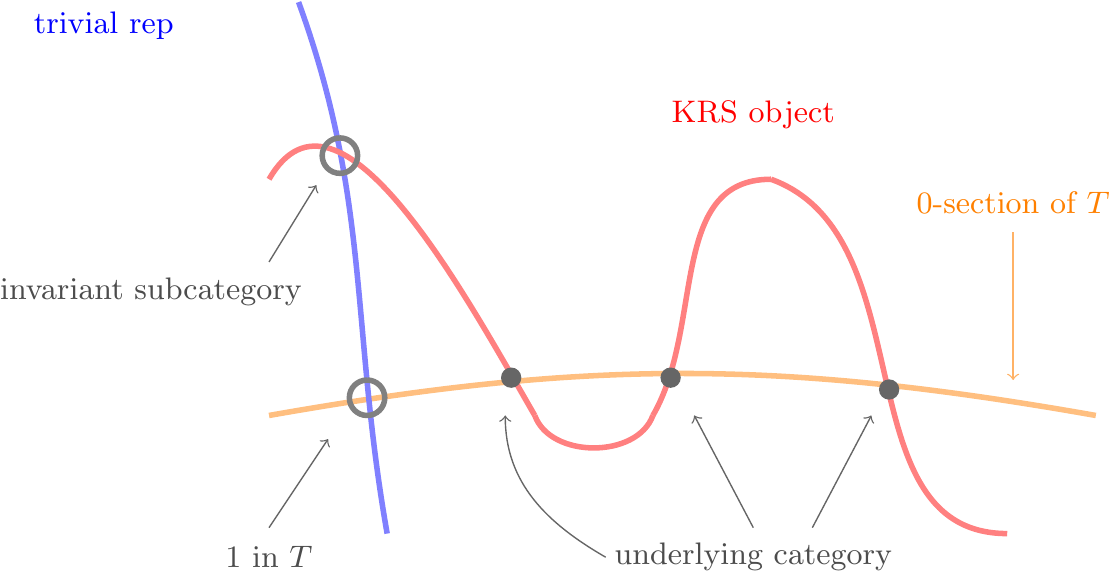} 
\includegraphics[scale=0.75]{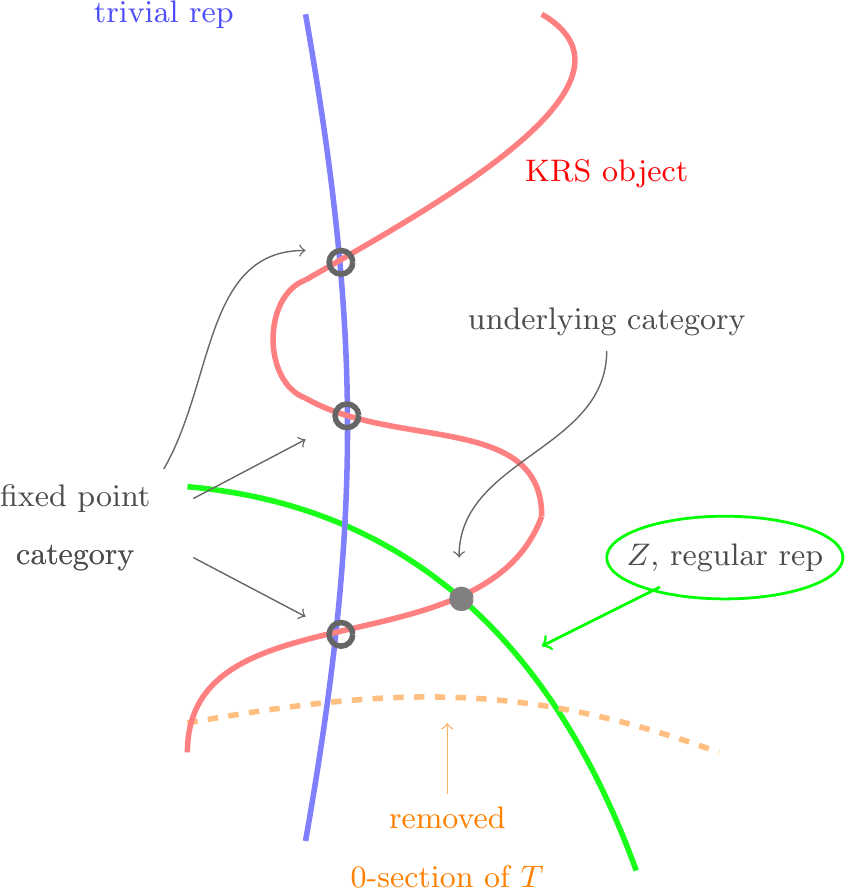} 
\caption{\small The BFM space for an abelian (LHS) and non-abelian (RHS) gauge theory, \cite{Teleman:2014jaa}. The KRS objects living in them are the objects in the 2-category referred to in the text and the trivial and regular representations are the Lagrangian submanifolds $L,L^{^{\prime}}$. Their intersection are Hom-spaces. These pictures generalise the LHS of figure \ref{fig:D}.}  
\label{fig:ctct}  
\end{center} 
\end{figure}

For the purpose of the present work we wish to highlight the role of the 2-fiber product used in \cite{Teleman:2014jaa} for identifying a preferred family of Lagrangian submanifolds foliating the BFM space, with the latter drawn in figure \ref{fig:ctct}, and therefore enabling to determine the Drinfeld center for cases exhibiting a generalised notion of homological mirror symmetry as the one addressed in section \ref{sec:DCFMQS}, \ref{sec:3.1} and \ref{sec:homms}.

\begin{figure}[ht!]      
\begin{center}
\includegraphics[scale=0.85]{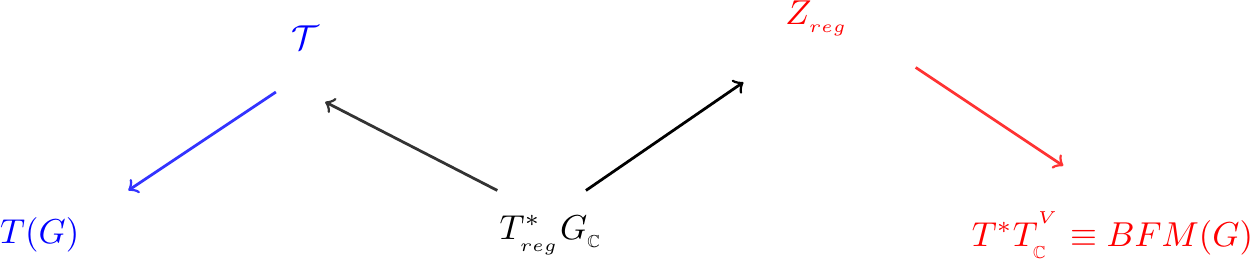}
\caption{\small The composition of fiber functors associated to double symplectic inductions leading to the regular and trivial representations. This diagram highlights the 2-fiber product structure shared with figure \ref{fig:a1aa2}.}
\label{fig:diagram}  
\end{center}
\end{figure}

Motivated by the realisation of homological mirror symmetry, the prescription of \cite{Teleman:2014jaa} makes use of a 2-fiber product, defined as follows

\begin{equation}
\boxed{\ \ \ {\cal T}\ \times_{_{T^{^*}_{_{reg}}G_{_{\mathbb{C}}}}}\ Z_{_{reg}}\ \ \color{white}\bigg] }, 
\label{eq:composite}
\end{equation}
with ${\cal T}$ and $Z_{_{reg}}$ denoting the trivial and regular fibers, respectively. \eqref{eq:composite} results from the composition outlined in figure \ref{fig:diagram}, where

\begin{equation}     
BFM(G)\ \overset{def.}{=}\ Z_{_{reg}}/G_{_{\mathbb{C}}}\ \ \ \ \ \ ,\ \ \ \ \ T(G)\ \overset{def.}{=}\ (N, \chi) \backslash\backslash T^*G_{_{\mathbb{C}}} // (N, \chi)\ \equiv\ N\ \backslash\ {\cal T}\ /N,
\end{equation}    
with

\begin{equation} 
\chi:\ n\ \longrightarrow\ \mathbb{C}^{^{\times}}
\end{equation}  
the regular character. Every such representation is symplectically induced from a 1D representation of a certain Levi subgroup $G$. As such, the preferred family of foliations resulting from \eqref{eq:composite} is the Fukaya category of a flag variety\footnote{A flag is a nested system of linear projective subspaces in a vector space. Given a field, $k$, a flag variety is the space of all flags in an $n$-dimensional $k$-vector space with the structure of a projective variety over $k$.} of $G$.

Given the considerations made in section \ref{sec:homms}, we therefore conclude that the aim of describing objects in the bulk of figure \ref{fig:ctct} is equivalent to that of defining the Drinfeld center for two absolute theories separated by a non-invertible defect, as encountered in \cite{Pasquarella:2023deo}. This further supports our main assertion, namely that, thanks to the prescription outlined in section \ref{sec:2.2}:

 \medskip    
   \medskip
\color{blue}

\noindent\fbox{%
    \parbox{\textwidth}{% 
   \medskip    
   \medskip
   \begin{minipage}{20pt}
        \ \ \ \ 
        \end{minipage}
        \begin{minipage}{380pt}
      \color{black}  In presence of multiple symplectically-induced representations, the resulting composite fiber functor defines the Drinfeld center of a mirror 3D ${\cal N}=4$ gauge theory, resulting from the union of multiple Coulomb branches of magnetic quivers associated to 3D ${\cal N}=4$ theories, thereby constituting a generalised realisation of homological mirror symmetry.
        \end{minipage}   
         \medskip    
   \medskip
        \\
    }%
}
 \medskip    
   \medskip     \color{black}

A more mathematical reformulation of this is the core topic of an upcoming work, \cite{VP1}. In particular, we aim to address how this correspondence can be exploited for treating the case where the algebraic varieties being dealt with are not complete intersections.

\section*{Key points} 

We can summarise the main results of the present work as follows: 

\begin{itemize} 

\item The 2-categorical description of the HB Hasse diagram of quiver gauge theories with 8 supercharges is equivalent to that of KRS. 
Importantly, this follows from the dimensional reduction of 4D ${\cal N}=2$ $SU(2)$ gauge theory with $N_{_f}=2$, as the most basic case with HB Hasse diagram consisting of two intersecting cones. 

\item Importantly, flatness of the moment map, ensuring the vanishing of higher homologies, \cite{DAlesio:2021hlp}, corresponds to the possibility of fully specifying the HS associated to the HB of a given theory. We emphasise that this is equivalent to stating the existence of a fiber functor in a 2-category, ultimately leading to the definition of a partition function of an absolute theory, according to the prescription of \cite{Freed:2022qnc}. 

\item The identification of the MQs whose CBs' union equals the HB of the original theory, corresponds to the identification of the Drinfeld center of a 3D theory, with the latter being a RW theory, $RW_{_X}$, whose BCs are Lagrangian submanifolds within a certain symplectic manifold, $X$. The complete reconstruction of the BFM space, where the objects of the 2-category live, corresponds to defining the Drinfeld center w.r.t. the given Lagrangian submanifolds and their intersection, i.e. the 1- and 2-morphisms characterising the 2-category in question. 

\item The fiber functor defining the partition function of interest\footnote{Namely the theory associated to the electric quiver in section \ref{sec:2.2}.} consists of a 2-fiber product, and is therefore compatible with the result of \cite{Teleman:2014jaa} in defining a preferred foliation for the BFM space, motivated by generalising homological mirror symmetry.

\end{itemize}

\section{Conclusions and outlook} 

Mostly inspired by \cite{Teleman:2014jaa}, this is the first of two papers by the same author addressing the formulation of mirror symmetry from the perspective of geometric representation theory. In this first work we propose a correspondence in between functorial field theory contructions, and Hasse diagrams resulting by implementing quiver subtraction on magnetic quivers (MQs)\footnote{With the latter being associated to the maximal decompositions of 5-brane webs on which a given absolute (electric) theory lives.}.

In section \ref{sec:DCFMQS} we reviewed the correspondence between geometric and algebraic resolutions of framed Nakajima quiver varieties, \cite{DAlesio:2021hlp}, highlighting it as an interesting example of homological mirror symmetry. In particular, we emphasised the property the moment map and higher homologies need to satisfy to ensure agreement in between the calculation of the two Hilbert series. We concluded the section with a brief overview of Hasse diagram construction via magnetic quivers for quiver gauge theories with 8 supercharges, pointing out an interesting 2-categorical structure when dealing with complete intersections. 

In section \ref{sec:3} we then turned to explaining how gauging-by-condensation can be related to the poset ordering leading to the construction of Hasse diagrams, thanks to the unifying role of the moment map. The identification of such moment map ensures the quiver gauge theory enjoys a generalised notion of homological mirror symmetry, with the latter corresponding to the presence of a Drinfeld center and a corresponding fiber functor for a 2-categorical structure, related to Rozansky-Witten theory, \cite{Rozansky:1996bq}. We concluded opening a connection between the topics outlined in the present work and those of \cite{Teleman:2014jaa, CT}, thereby setting the stage for a more mathematical treatment to which \cite{VP1} is devoted.

\subsection*{Acknowlegements} 

I wish to thank Tudor Dimofte, Fernando Quevedo,  Marcus Sperling and Constantin Teleman, for very useful discussions and questions raised at different stages of the present work. 
This work is partially supported by an STFC scholarship through DAMTP.

\appendix


\begin{thebibliography}{99}   


\bibitem{Teleman:2014jaa}
C.~Teleman,
\emph{Gauge theory and mirror symmetry},  
[arXiv:1404.6305 [math-ph]].



\bibitem{CT}
C.~Teleman,
to appear.


\bibitem{Pasquarella:2023deo}
V.~Pasquarella,
\emph{Categorical Symmetries and Fiber Functors from Multiple Condensing Homomorphisms from 6D ${\cal N}=(2,0)$ SCFTs},
[arXiv:2305.18515 [hep-th]].

\bibitem{Freed:2022qnc}
D.~S.~Freed, G.~W.~Moore and C.~Teleman,
\emph{Topological symmetry in quantum field theory},   
[arXiv:2209.07471 [hep-th]]. 

\bibitem{VP1}
V.~Pasquarella,
to appear.  



\bibitem{Cabrera:2019izd}
S.~Cabrera, A.~Hanany and M.~Sperling,
\emph{Magnetic quivers, Higgs branches, and 6d $N$=(1,0) theories},   
JHEP \textbf{06} (2019), 071
[erratum: JHEP \textbf{07} (2019), 137]
doi:10.1007/JHEP06(2019)071
[arXiv:1904.12293 [hep-th]].


\bibitem{Bourget:2019aer}
A.~Bourget, S.~Cabrera, J.~F.~Grimminger, A.~Hanany, M.~Sperling, A.~Zajac and Z.~Zhong,
\emph{The Higgs mechanism \textemdash{} Hasse diagrams for symplectic singularities}, 
JHEP \textbf{01} (2020), 157
doi:10.1007/JHEP01(2020)157
[arXiv:1908.04245 [hep-th]].

\bibitem{Bourget:2021siw}
A.~Bourget, J.~F.~Grimminger, A.~Hanany, M.~Sperling and Z.~Zhong,
\emph{Branes, Quivers, and the Affine Grassmannian}, 
Adv. Stud. Pure Math. \textbf{88} (2023), 331-435
doi:10.2969/aspm/08810331
[arXiv:2102.06190 [hep-th]].


\bibitem{Bourget:2023cgs}
A.~Bourget, J.~F.~Grimminger, A.~Hanany, R.~Kalveks, M.~Sperling and Z.~Zhong,
\emph{A Tale of N Cones},   
[arXiv:2303.16939 [hep-th]].

\bibitem{Bourget:2019rtl}
A.~Bourget, S.~Cabrera, J.~F.~Grimminger, A.~Hanany and Z.~Zhong,
\emph{Brane Webs and Magnetic Quivers for SQCD},  
JHEP \textbf{03} (2020), 176
doi:10.1007/JHEP03(2020)176
[arXiv:1909.00667 [hep-th]].


\bibitem{Ferlito:2016grh}
G.~Ferlito and A.~Hanany,
\emph{A tale of two cones: the Higgs Branch of Sp(n) theories with 2n flavours},    
[arXiv:1609.06724 [hep-th]]. 

\bibitem{Hanany:2019tji}
A.~Hanany and R.~Kalveks,
\emph{Quiver Theories and Hilbert Series of Classical Slodowy Intersections},  
Nucl. Phys. B \textbf{952} (2020), 114939
doi:10.1016/j.nuclphysb.2020.114939
[arXiv:1909.12793 [hep-th]].

\bibitem{Rozansky:1996bq}
L.~Rozansky and E.~Witten,
\emph{HyperKahler geometry and invariants of three manifolds},
Selecta Math. \textbf{3} (1997), 401-458
doi:10.1007/s000290050016
[arXiv:hep-th/9612216 [hep-th]].


\bibitem{DAlesio:2021hlp}
S.~D\textquoteright{}Alesio,
\emph{Derived representation schemes and Nakajima quiver varieties},   
Selecta Math. \textbf{28} (2022) no.1, 20
doi:10.1007/s00029-021-00724-4  

\bibitem{Braverman:2016pwk}
A.~Braverman, M.~Finkelberg and H.~Nakajima,
\emph{Coulomb branches of $3d$ $\mathcal{N}=4$ quiver gauge theories and slices in the affine Grassmannian}, 
Adv. Theor. Math. Phys. \textbf{23} (2019), 75-166
doi:10.4310/ATMP.2019.v23.n1.a3
[arXiv:1604.03625 [math.RT]].


\bibitem{Braverman:2016wma}
A.~Braverman, M.~Finkelberg and H.~Nakajima,
\emph{Towards a mathematical definition of Coulomb branches of $3$-dimensional $\mathcal{N} = 4$ gauge theories, II},  
Adv. Theor. Math. Phys. \textbf{22} (2018), 1071-1147
doi:10.4310/ATMP.2018.v22.n5.a1
[arXiv:1601.03586 [math.RT]].




\bibitem{Atiyah:1978ri}
M.~F.~Atiyah, N.~J.~Hitchin, V.~G.~Drinfeld and Y.~I.~Manin,
\emph{Construction of Instantons}, 
Phys. Lett. A \textbf{65} (1978), 185-187
doi:10.1016/0375-9601(78)90141-X.


\bibitem{Berest}  
Y.~Berest, G.~Khachatryan, A.~Ramadoss, \emph{Derived representation schemes and cyclic homology}, Adv.
Math. 245, 625–689 (2013).


\bibitem{Cabrera:2018ann}
S.~Cabrera and A.~Hanany,
\emph{Quiver Subtractions},  
JHEP \textbf{09} (2018), 008
doi:10.1007/JHEP09(2018)008
[arXiv:1803.11205 [hep-th]].


\bibitem{Teleman:2018wac}
C.~Teleman,
\emph{The r\^ole of Coulomb branches in 2D gauge theory}, 
J. Eur. Math. Soc. \textbf{23} (2021) no.11, 3497-3520
doi:10.4171/jems/1071
[arXiv:1801.10124 [math.AG]].   

\bibitem{GIT}
C.~Teleman. \emph{The Quantization Conjecture Revisited.}, Annals of Mathematics 152, no. 1 (2000): 1–43. https://doi.org/10.2307/2661378.

\bibitem{Gonzalez:2023jur}
E.~Gonzalez, C.~Y.~Mak and D.~Pomerleano,
\emph{Coulomb branch algebras via symplectic cohomology},   
[arXiv:2305.04387 [math.SG]].




 

\bibitem{Seiberg:1996nz}
N.~Seiberg and E.~Witten,
\emph{Gauge dynamics and compactification to three-dimensions},
[arXiv:hep-th/9607163 [hep-th]].  



\bibitem{Kapustin:2009uw}
A.~Kapustin and L.~Rozansky,
\emph{Three-dimensional topological field theory and symplectic algebraic geometry II},
Commun. Num. Theor. Phys. \textbf{4} (2010), 463-550
doi:10.4310/CNTP.2010.v4.n3.a1
[arXiv:0909.3643 [math.AG]].

\bibitem{Kapustin:2008sc}
A.~Kapustin, L.~Rozansky and N.~Saulina,
\emph{Three-dimensional topological field theory and symplectic algebraic geometry I},
Nucl. Phys. B \textbf{816} (2009), 295-355
doi:10.1016/j.nuclphysb.2009.01.027
[arXiv:0810.5415 [hep-th]].


\bibitem{BFM}
R.~Bezrukavnikov, M.~Finkelberg, and I.~Mirkovíc, \emph{Equivariant homology and K-theory of affine Grassmannians and Toda lattices},   
Compos.Math. 141 (2005) 746-768.



\bibitem{Bhardwaj:2023zix}
L.~Bhardwaj, M.~Bullimore, A.~E.~V.~Ferrari and S.~Schafer-Nameki,
\emph{Generalized Symmetries and Anomalies of 3d N=4 SCFTs},
[arXiv:2301.02249 [hep-th]].   


\bibitem{Gaiotto:2020iye}
D.~Gaiotto and J.~Kulp,
\emph{Orbifold groupoids}, 
JHEP \textbf{02} (2021), 132
doi:10.1007/JHEP02(2021)132
[arXiv:2008.05960 [hep-th]].


\bibitem{Kong:2020cie}
L.~Kong, T.~Lan, X.~G.~Wen, Z.~H.~Zhang and H.~Zheng,
\emph{Algebraic higher symmetry and categorical symmetry -- a holographic and entanglement view of symmetry},   
Phys. Rev. Res. \textbf{2} (2020) no.4, 043086
doi:10.1103/PhysRevResearch.2.043086
[arXiv:2005.14178 [cond-mat.str-el]].

\bibitem{Kong:2022cpy}
L.~Kong and Z.~H.~Zhang,
\emph{An invitation to topological orders and category theory}, 
[arXiv:2205.05565 [cond-mat.str-el]].

\bibitem{Kong:2019byq}
L.~Kong and H.~Zheng,
\emph{A mathematical theory of gapless edges of 2d topological orders. Part I}, 
JHEP \textbf{02} (2020), 150
doi:10.1007/JHEP02(2020)150
[arXiv:1905.04924 [cond-mat.str-el]].

\bibitem{Kong:2019cuu}
L.~Kong and H.~Zheng,
\emph{A mathematical theory of gapless edges of 2d topological orders. Part II},  
Nucl. Phys. B \textbf{966} (2021), 115384
doi:10.1016/j.nuclphysb.2021.115384
[arXiv:1912.01760 [cond-mat.str-el]].


\bibitem{Kong:lastbutone}
L.~Kong and H.~Zheng,
\emph{Categorical computation}, 
Front. Phys. 18(2), 21302 (2023)
[arXiv:2102.04814v2[quant-ph]].



\bibitem{Gaiotto:2019xmp}
D.~Gaiotto and T.~Johnson-Freyd,
\emph{Condensations in higher categories}, 
[arXiv:1905.09566 [math.CT]].

\bibitem{Johnson-Freyd:2021tbq}
T.~Johnson-Freyd and M.~Yu,
\emph{Topological Orders in (4+1)-Dimensions},   
SciPost Phys. \textbf{13} (2022) no.3, 068
doi:10.21468/SciPostPhys.13.3.068
[arXiv:2104.04534 [hep-th]].



\bibitem{Johnson-Freyd:2020usu}
T.~Johnson-Freyd,
\emph{On the Classification of Topological Orders},   
Commun. Math. Phys. \textbf{393} (2022) no.2, 989-1033
doi:10.1007/s00220-022-04380-3
[arXiv:2003.06663 [math.CT]].


\bibitem{MYu}
T.~D.~D\'ecoppet and M.~Yu,
\emph{Gauging noninvertible defects: a 2-categorical perspective},
Lett. Math. Phys. \textbf{113} (2023) no.2, 36
doi:10.1007/s11005-023-01655-1
[arXiv:2211.08436 [math.CT]].




\bibitem{Freed:2012bs}
D.~S.~Freed and C.~Teleman,
\emph{Relative quantum field theory},    
Commun. Math. Phys. \textbf{326} (2014), 459-476
doi:10.1007/s00220-013-1880-1
[arXiv:1212.1692 [hep-th]].






    

\bibitem{TJF}    
T.~Johnson-Freyd,    
\emph{Operators and higher categories in quantum field theory}, Lecture series. 

\bibitem{Hori:2000kt}
K.~Hori and C.~Vafa,
\emph{Mirror symmetry},  
[arXiv:hep-th/0002222 [hep-th]].

\bibitem{Klemm:1996bj}
A.~Klemm, W.~Lerche, P.~Mayr, C.~Vafa and N.~P.~Warner,
\emph{Selfdual strings and N=2 supersymmetric field theory},  
Nucl. Phys. B \textbf{477} (1996), 746-766
doi:10.1016/0550-3213(96)00353-7
[arXiv:hep-th/9604034 [hep-th]].

\bibitem{Kontsevich:1994dn}
M.~Kontsevich,
\emph{Homological Algebra of Mirror Symmetry},  
[arXiv:alg-geom/9411018 [math.AG]].

\bibitem{Gaiotto:2008ak}
D.~Gaiotto and E.~Witten,
\emph{S-Duality of Boundary Conditions In N=4 Super Yang-Mills Theory},   
Adv. Theor. Math. Phys. \textbf{13} (2009) no.3, 721-896
doi:10.4310/ATMP.2009.v13.n3.a5
[arXiv:0807.3720 [hep-th]].


\bibitem{CT1}
C.~Teleman,
online talk.  


\bibitem{mommap}
W. Crawley-Boevey, Geometry of the moment map for representations of quivers, Compositio Math. 126 (2001), no. 3,
257–293.



\bibitem{KP}
H. Kraft, C. Procesi, 
\emph{Minimal singularities inGL n},
 Invent Math 62, 503–515 (1980). https://doi.org/10.1007/BF01394257.

\bibitem{KP1}  
H. Kraft, C. Procesi,  
\emph{On the geometry of conjugacy classes in classical groups},
Comment.Math.Helv. 57 (1982) 539 • DOI: 10.1007/BF02565876.

\end{thebibliography}
\end{document}